\documentclass[aps,prb,twocolumn,floatfix,groupedaddress]{revtex4-2}
\usepackage{graphicx,amsmath,amsfonts,mathrsfs,bbold,dsfont}
\begin{document}

\title{Basic aspects of ferroelectricity induced by noncollinear alignment of spins}

\author{I.~V.~Solovyev}
\email{SOLOVYEV.Igor@nims.go.jp}
\affiliation{Research Center for Materials Nanoarchitectonics (MANA), National Institute for Materials Science (NIMS), 1-1 Namiki, Tsukuba, Ibaraki 305-0044, Japan}

\date{\today}

\date{\today}
\begin{abstract}
Basic principles of ferroelectric activity induced by the noncollinear alignment of spins are reviewed. First, there is a fundamental reason why the inversion symmetry can be broken by certain magnetic order. Such situation occurs when the magnetic order simultaneously involves ferromagnetic ($\boldsymbol{F}$) and antiferromagnetic ($\boldsymbol{A}$) patterns, transforming under the spatial inversion $\mathcal{I}$ and time reversal $\mathcal{T}$ as $\mathcal{I}\boldsymbol{F}=\boldsymbol{F}$ and $\mathcal{IT}\boldsymbol{A}=\boldsymbol{A}$, respectively. The incompatibility of these two conditions results in breaking the inversion symmetry and appearance of electric polarization $\vec{P}$. The noncollinear alignment of spins is just one of examples of such coexistence of $\boldsymbol{F}$ and $\boldsymbol{A}$. This coexistence principle imposes a constraint on possible dependencies of $\vec{P}$ on the directions of spins, which can include only two types of terms: ``antisymmetric coupling'' and ``single-ion anisotropy'' in the from $\vec{P} = \vec{\boldsymbol{\mathcal{P}}}_{12} \cdot [ \boldsymbol{e}_{1} \times \boldsymbol{e}_{2} ] + \boldsymbol{e}_{1} \cdot \vec{\mathbb{\Pi}} \boldsymbol{e}_{1} - \boldsymbol{e}_{2} \cdot \vec{\mathbb{\Pi}} \boldsymbol{e}_{2}$. Microscopically, $\vec{\boldsymbol{\mathcal{P}}}_{12}$ can be evaluated in the framework of superexchange theory. For the single Kramers doublet, this theory yields $\vec{\boldsymbol{\mathcal{P}}}_{12} \sim t_{12}^{0} \vec{\boldsymbol{r}}_{12}^{\phantom{0}}$, where $t_{12}^{0}$ is the spinless component of transfer integrals and $\vec{\boldsymbol{r}}_{12}^{\phantom{0}}$ is the spin-dependent part of position operator induced by the relativistic spin-orbit coupling. The construction $t_{12}^{0} \vec{\boldsymbol{r}}_{12}^{\phantom{0}}$ remains invariant under spatial inversion, providing a clear microscopic reason why noncollinear alignment of spins can induce $\vec{P}$ even in centrosymmetric crystals. Then, the symmetry properties of $\vec{\boldsymbol{r}}_{12}^{\phantom{0}}$ can be rationalized from the viewpoint of symmetry of Kramers states. Particularly, the commonly used Katsura-Nagaosa-Balatsky (KNB) rule $\vec{P} \propto \vec{\epsilon}_{21} \times [\boldsymbol{e}_{1} \times \boldsymbol{e}_{2}]$ ($\vec{\epsilon}_{21}$ being the bond direction) can be justified only for relatively high symmetry of the bond (the threefold rotational symmetry or higher). Furthermore, the single-ion anisotropy $\vec{\mathbb{\Pi}}$ should vanish for the spin $\frac{1}{2}$ or if magnetic ions are located in inversion centers, thus severely restricting the applicability of this microscopic mechanism. The properties of multiferroic materials are reconsidered from the viewpoint of these principles. A particular attention is paid to complications caused by possible deviations from the KNB rule.  
\end{abstract}

\maketitle

\section{\label{sec:Intro} Introduction}
\par The ferroelectricity is the property of insulating materials to possess spontaneous electric polarization $\vec{P}$~\cite{LinesGlass}. Since the spatial inversion operation $\mathcal{I}$ changes the sign of $\vec{P}$, it is essential that $\mathcal{I}$ must be broken. In this sense, the search of new ferroelectric materials is basically the search of new mechanisms of breaking the inversion symmetry and possibilities to realize these mechanisms in practice. 

\par Canonically, the inversion symmetry breaking is associated with the properties of nonmagnetic crystalline lattice, whose atomic positions do not transform to themselves by $\mathcal{I}$. Here, the polar distortion induces the additional mixing between occupied bonding and unoccupied antibonding states and, if this mixing lowers the energy, the distortion becomes stable~\cite{Bersuker1966,Cohen1992}. The typical mechanism promoting such mixing is the transition metal $3d$ - oxygen $2p$ hybridization in $d^{\,0}$ perovskite oxides like PbTiO$_3$ and BaTiO$_3$~\cite{Cohen1992}.

\par A very special case of ferroelectricity is when $\mathcal{I}$ is broken by a magnetic alignment of spins. From the canonical point of view, the idea may look rather counterintuitive, because the magnetism is typically associated with breaking time reversal symmetry $\mathcal{T}$, while breaking $\mathcal{I}$ was believed to be the prerogative of atomic displacements~\cite{Hill}. Nevertheless, such a possibility was predicted in 1960 by Dzyaloshinskii~\cite{DzyaloshinskiiME}. He realized that the magnetic structure of some antiferromagnetic (AFM) materials, such as Cr$_2$O$_3$, can be transformed to itself by combining $\mathcal{I}$ with $\mathcal{T}$. Therefore, both $\mathcal{I}$ and $\mathcal{T}$ can be simultaneously broken by applying a magnetic field $\boldsymbol{H}$, which will induce not only the net magnetization, but also the electric polarization $\vec{P}$. Such materials are now called the magnetoelectrics. The key point here that $\boldsymbol{H}$ deforms the AFM alignment of spins and make it noncollinear, which is crucial for breaking the inversion symmetry. Therefore, the next step was to realize the same effect but without external magnetic field, by exploiting for these purposes the intrinsic noncollinearity of spin-spiral magnets arising from the competition of several exchange interactions. The materials, where the inversion symmetry is spontaneously broken by intrinsic magnetic order without external magnetic field are called type-II multiferroics. We do not consider here type-I multiferroics, where the ferroelectricity and ferromagnetism have a different origin~\cite{Hill,Khomskii2009}. Therefore, ``multiferroics'' in this review will solely mean type-II multiferroics. This direction became extremely popular in early 2000s, after the discovery of ferroelectric activity in the spin-spiral magnet TbMnO$_3$~\cite{Kimura_TbMnO3} and other materials~\cite{CheongMostovoy}, and still attracts a lot of attention today as it provides a unique possibility for the cross-control of magnetization and polarization by electric and magnetic fields~\cite{Eerenstein,Tokura}. Nevertheless, it is worth to remark that although the primary emphasize was given to the search of spin-spiral multiferroics~\cite{CheongMostovoy,TokuraSeki}, historically they were not the first multiferroics. For instance, the inversion symmetry breaking arising from incompatibility of magnetic orders in the rare-earth ($R$) and transition-metal ($M$) sublattices was predicted in orthorhombically distorted perovskites $RM$O$_3$ as early as in 1973~\cite{Yamaguchi}.

\par The modern theory of electric polarization was developed by King-Smith and Vanderbilt in early 1990s~\cite{FE_theory1,FE_theory2}. It has two equivalent formulations: in the reciprocal space, in terms of the Berry connection, and in the real space, in terms of the Wannier functions $| w_{i} \rangle$ for the occupied bands. The latter formulation is more convenient for the construction of microscopic models, using for these purposes the perturbation theory in the real space. It states that the electric polarization $\vec{P}$ (actually, the change of polarization in the process of adiabatic lowering the symmetry) is given by expectation values of the position operator $\vec{r}$ in the basis of these Wannier functions as
\noindent
\begin{equation}
\vec{P}=-\frac{e}{V}\sum\limits_{i} \langle w_{i} | \vec{r} \, | w_{i} \rangle,
\label{eq:elpol}
\end{equation}
\noindent where $-$$e$ is the electron charge, $V$ is the volume, and $i$ is the lattice point (the Wannier center)~\cite{FE_theory1,FE_theory2,FE_theory3}. This theory was spurred by growing interest in ferroelectric materials as well as the phenomenon of ferroelectricity itself and provided a long-awaited solution of the problem on how to properly calculate $\vec{P}$ in periodic systems, especially in first-principles electronic structure methods. It was still one decade before the new era of magnetic multiferroics, sparkled in early 2000s, and no consideration of magnetic dependencies of electric polarization was given around that time. Nevertheless, the idea seems to be straightforward: since $\vec{r}$ does not depend on magnetic degrees of freedom, all information about the magnetism is included in $| w_{i} \rangle$ and what we have to do is to understand how $| w_{i} \rangle$ evolves with the change of magnetic structures.

\par The Katsura-Nagaosa-Balatsky (KNB) rule for electric polarization induced by spiral magnetic order was formulated in 2005~\cite{KNB}, soon after discovery of this effect in TbMnO$_3$~\cite{Kimura_TbMnO3}. It relates the direction of polarization in the bond with the ones of magnetic moments $\boldsymbol{e}_{1}$ and $\boldsymbol{e}_{2}$ as $\vec{P} \propto \vec{\epsilon}_{21} \times [\boldsymbol{e}_{1} \times \boldsymbol{e}_{2}]$, where $\vec{\epsilon}_{21}$ is the direction of the bond. The rule appeared to be rather universal and readily explained the main trends in the behavior of quite a large number of multiferroic materials with spiral magnetic order~\cite{Khomskii2009,CheongMostovoy,TokuraSeki,TokuraSekiNagaosa}.   

\par The KNB theory, being the first microscopic theory of magnetoelectric coupling and explaining the appearance of electric polarization in noncollinear magnets, played a very important role in understanding the properties of multiferroic materials and the phenomenon of multiferroicity itself. However, there are also many concerns about universality, validity, and generality of the KNB rule. In their original work, KNB considered a very simple and very special model for the centrosymmetric bond, in which the magnetic $d$ states at the edges were mediated by the ligand $p$ states in the midpoint of the bond. The magnetic states were also taken in the very special form of twofold degenerate $\Gamma_{7}$ representation formed by cubic $xy$, $yz$, and $zx$ orbitals in the presence of relativistic spin-orbit (SO) coupling. Thus, it is not clear to which extend the expression for the electric polarization, derived for this special case, can be and should be applied for the analysis of magnetoelectric coupling in all types of multiferroic materials, as it is typically done today. Moreover, despite a formal simplicity, the KNB model does not provided a clear answer why the noncollinear magnetic order breaks the inversion symmetry and which microscopic invariant is responsible for finite polarization in centrosymmetric crystals.

\par The situation was further complicated by discoveries of new multiferroic materials, where the electric polarization is induced by proper screw magnetic order with the spins rotating in the plane perpendicular to the bonds, i.e. where the canonical KNB theory predicts no electric polarization~\cite{Arima,RbFeMo2O8,MnI2}. This has led to the proposal of yet another popular mechanism of magnetically induced ferroelectricity, the so-called spin-dependent metal-ligand hybridization mechanism~\cite{Arima}, which is regarded as supplementary to the KNB one~\cite{TokuraSekiNagaosa}. However, even though the conventional KNB picture may look at odds with the properties of certain multiferroic materials, is it because of additional simplifications considered by KNB (while the basic idea itself is still vital) or does it mean that this mechanism cannot work in principle and one have to consider alternative possibilities? 

\par The goal of this review article is to explain, from the very basic point of view, why and how the electric polarization can be induced by the noncollinear alignment of spins. We start with simple phenomenological considerations and argue why the noncollinear alignment of spins breaks the inversion symmetry. This phenomenological picture will help us to understand how the electric polarization should depend on the relative directions of spins. Particularly, for the centrosymmetric bond, there can be only two contributions to $\vec{P}$: in the form of the antisymmetric exchange coupling and single-ion anisotropy. 

\par Then, we turn to a microscopic picture for the magnetically dependent polarization. Formally, the derivation of KNB does not rely on modern theory of electric polarization. However, the perturbation theory considered by KNB can be viewed as a particular way to evaluate Wannier functions $| w_{i} \rangle$ for the occupied states, which is required by the modern theory. We will explain how this can be done rigorously by considering the superexchange-type theory for electric polarization. This theory will give us a clear answer which microscopic parameter is responsible for the magnetoelectric coupling in centrosymmetric materials. The form of the coupling between noncollinear spins and the electric polarization can be more general than the one prescribed by the KNB rule. The latter can be recovered only for threefold rotational or higher symmetries of the bond. Furthermore, we will rationalize the behavior of the single-ion anisotropy for the electric polarization and argue that this contribution should vanish at least in the following two cases: for the spin $\frac{1}{2}$ and if the magnetic ion is located in the centrosymmetric position. Since the spin-dependent metal-ligand hybridization is just one of the microscopic mechanism for the single-ion anisotropy, our finding severely restricts its applicability. With this in mind we will reconsider the behavior of known multiferroic materials and the roles played by different microscopic mechanisms in this behavior. 

\section{\label{sec:ireason} Reasons of inversion symmetry breaking}
\par
\begin{figure}[b]
\begin{center}
\includegraphics[width=8.6cm]{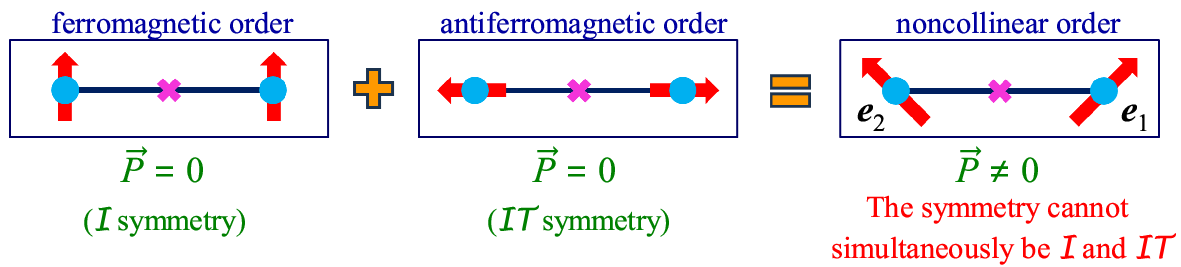}
\end{center}
\caption{
Cartoon picture explaining inversion symmetry breaking by noncollinear magnetic alignment in otherwise centrosymmetric bond. The crystallographic inversion center is denoted by $\times$. The ferromagnetic order ($\boldsymbol{F}$) is invariant under the spatial inversion $\mathcal{I}$. The antiferromagnetic order ($\boldsymbol{A}$) is invariant under the symmetry operation $\mathcal{IT}$, where $\mathcal{I}$ is combined with the time reversal $\mathcal{T}$. In both cases, the electric polarization $\vec{P}$ is equal to zero. The noncollinear order can be viewed as a superposition of the collinear orders $\boldsymbol{F}$ and $\boldsymbol{A}$. Since the $\mathcal{I}$ and $\mathcal{IT}$ symmetries (required by, respectively, $\boldsymbol{F}$ and $\boldsymbol{A}$ counterparts) cannot coexist, the inversion symmetry is broken, resulting in finite $\vec{P}$. $\boldsymbol{e}_{1}$ and $\boldsymbol{e}_{2}$ are the directions of spins at two sites of the bonds.}
\label{fig:ireason}
\end{figure}

 Consider a bond connecting the sites $1$ and $2$ with the inversion center in the midpoint (Fig.~\ref{fig:ireason}). Let $\boldsymbol{e}_{1}$ and $\boldsymbol{e}_{2}$ be the directions of spins at the sites $1$ and $2$, which is normalized to unity. Then, any noncollinear alignment of spins $\boldsymbol{e}_{1}$ and $\boldsymbol{e}_{2}$ can be viewed as a superposition of ferromagnetic (FM) and AFM order parameters, which are given by, respectively, $\boldsymbol{F} = \boldsymbol{e}_{1} + \boldsymbol{e}_{2}$ and $\boldsymbol{A} = \boldsymbol{e}_{1} - \boldsymbol{e}_{2}$. $\boldsymbol{F}$ has the symmetry of the bond and is transformed to itself by the spatial inversion about the midpoint: $\mathcal{I} \boldsymbol{F} = \boldsymbol{F}$. However, in order to keep $\boldsymbol{A}$ invariant, it is essential to combine $\mathcal{I}$ with $\mathcal{T}$, which additionally transforms each $\boldsymbol{e}_{i}$ to $-\boldsymbol{e}_{i}$ ($i=$ $1$ or $2$), yielding $\mathcal{IT} \boldsymbol{A} = \boldsymbol{A}$.
\noindent
\noindent Thus, although inversion symmetry can be preserved separately for the $\boldsymbol{F}$ or $\boldsymbol{A}$ order (either as itself or in the combination with $\mathcal{T}$), it appears to be broken in the case of noncollinear alignment of spins, where both order parameters are finite: simply, there is no such symmetry operation, which would simultaneously transform $\boldsymbol{F}$ and $\boldsymbol{A}$ to themselves. Hence, the system can develop finite polarization $\vec{P}$. The explanation is general and equally apply to the canonical magnetoelectric effect in antiferromagnetic Cr$_2$O$_3$~\cite{DzyaloshinskiiME}, where the noncollinearity of spins is induced by $\boldsymbol{H}$, destroying the $\mathcal{IT}$ symmetry of AFM alignment, or improper multiferroics, where this noncollinearity stems from competition of isotropic exchange interactions~\cite{KNB}. In this sense, the multiferroic can be viewed as an intrinsic magnetoelectric, in which the internal field due to $\boldsymbol{F}$ plays the role of $\boldsymbol{H}$.

\par In the notations here and elsewhere in the paper, the vector symbol is reserved for the direction of electric polarization $\vec{P}$ and related to it quantities such as the position operator $\vec{r}$ and electric field $\vec{E}$, while other vector quantities, related to magnetic degrees of freedom, are denoted by bold symbols. 

\section{\label{sec:phenomenology} Spin dependence of induced polarization: \\ a phenomenological view}
\par In the previous section, we have seen that in order to break the inversion symmetry by magnetic means, the ferromagnetic order parameter in the centrosymmetric bond, $\boldsymbol{F}$, should coexist with the antiferromagnetic one, $\boldsymbol{A}$, meaning that the spins in the bond are noncollinear to each other. Therefore, the induced polarization $\vec{P}$ should be proportional to both $\boldsymbol{F}$ and $\boldsymbol{A}$. The most general combination of $\boldsymbol{F}$ and $\boldsymbol{A}$ is the $3$$\times$$3$ matrix $\boldsymbol{F}^{T} \otimes \boldsymbol{A}$ of Kronecker product, where $\boldsymbol{F}^{T}$ is the row-vector, corresponding to the column-vector $\boldsymbol{F}$. Then, it is straightforward to find (see Supplemental Material~\cite{SM}) that:
\noindent
\begin{equation}
\vec{P} = \vec{\boldsymbol{\mathcal{P}}}_{12} \cdot [ \boldsymbol{e}_{1} \times \boldsymbol{e}_{2} ] + \boldsymbol{e}_{1} \cdot \vec{\mathbb{\Pi}} \boldsymbol{e}_{1} - \boldsymbol{e}_{2} \cdot \vec{\mathbb{\Pi}} \boldsymbol{e}_{2} .
\label{eq:pol}
\end{equation}
\noindent Thus, there can be only two source of polarization in centrosymmetric bonds. The first one is the antisymmetric coupling, which is proportional to the cross product of $\boldsymbol{e}_{1}$ and $\boldsymbol{e}_{2}$, similar to antisymmetric Dzyaloshinskii-Moriya (DM) interaction $\boldsymbol{D}_{12} \cdot [ \boldsymbol{e}_{1} \times \boldsymbol{e}_{2} ]$~\cite{Dzyaloshinskii_weakF,Moriya_weakF}. Nevertheless, $\boldsymbol{D}_{12} = [D_{12}^{c}] $ is a vector, while $\vec{\boldsymbol{\mathcal{P}}}_{12} = [ \mathcal{P}_{12}^{v,c} ]$ is the rank-2 tensor, where the first index ($v$) is reserved for the vector component of electric polarization and the second index ($c$) stands to describe the coupling with the spin vector, which in this case is given by the cross product $[ \boldsymbol{e}_{1} \times \boldsymbol{e}_{2} ]$. Furthermore, $\boldsymbol{D}_{12}$ vanishes in the centrosymmetric bond, while $\vec{\boldsymbol{\mathcal{P}}}_{12}$ does not. The second source is single-ion anisotropy, described by the rank-3 tensor $\vec{\mathbb{\Pi}} = [\Pi^{v,ab}]$ (where $v$ is referred to the direction of electric polarization, while $a$ and $b$ describe the coupling with spins). If the bond is centrosymmetric, the single-ion contribution is finite only when $\boldsymbol{e}_{2} \ne \boldsymbol{e}_{1}$ (i.e., the spins are noncollinear).

\par No other contributions to $\vec{P}$ except these two are expected. For instance, there should be no symmetric (with respect to the permutation of $\boldsymbol{e}_{1}$ and $\boldsymbol{e}_{2}$) contributions to $\vec{P}$, neither isotropic nor anisotropic ones. If $1$ and $2$ are the edges of isolated dimer, the tensor $\vec{\mathbb{\Pi}}$ is finite (except for the special case of $S=\frac{1}{2}$~\cite{PRB2020}, which will be considered in Sec.~\ref{sec:s12}). Nevertheless, if the bond is a part of the solid, it is quite common that the atoms forming the bond are located in inversion centers. Then, $\vec{\mathbb{\Pi}}$ should vanish.

\par According to these phenomenological considerations, most of microscopic models, so far proposed for the analysis of magnetoelectric coupling in noncollinear magnets, were focusing on either the antisymmetric exchange in magnetic bonds~\cite{KNB,PRB2017} or the single-ion anisotropy~\cite{Arima,PRB2015b}. Other forms of magnetic dependence of the electric polarization in noncollinear magnets would be exotic, at least from this phenomenological point of view. 

\section{\label{sec:microscopic} Microscopic picture}
\par The simplest microscopic expression for $\vec{\boldsymbol{\mathcal{P}}}_{12}$ can be derived by considering the following chain of arguments~\cite{PRL2021}. Since the electric polarization in metals is screened by free electrons, the ferroelectricity is the property of insulating materials. Then, vast majority of magnetic insulators are Mott insulators, where both magnetism and insulating behavior are due to the large on-site Coulomb repulsion $U$. Therefore, it is natural to assume that magnetic ferroelectrics (or multiferroics) belong to the same category. The key theory for interatomic exchange interactions in Mott insulators is Anderson's superexchange (SE) theory~\cite{Anderson1959}. It starts with the limit of large $U$ and in each bond ($ij$) treats electronic hoppings $\hat{t}_{ij}$ as a perturbation. Basically, the SE theory is the second order perturbation theory for the energy with respect to $\hat{t}_{ij}$. Our goal is to formulate a similar theory for the polarization $\vec{P}$ given by Eq.~\ref{eq:elpol}. Since the position operator itself does not depend on spin, the magnetoelectric coupling arises entirely from the spin dependence of the Wannier functions. Then, the first observation is that, instead of the second order perturbation theory for the energy, the SE theory can be formulated by considering the first order perturbation theory for $| w_{i} \rangle$. Indeed, if $| w_{i} \rangle = | \alpha_{i}^{o} \rangle$ is the occupied atomic state in the limit $U \to \infty$, the electronic hoppings will induce a tail spreading to the neighboring site $j$:
\noindent
\begin{equation}
| w_{i} \rangle \approx  | \alpha_{i}^{o} \rangle + | \alpha_{i \to j}^{o} \rangle.
\label{eq:wfocc}
\end{equation}
\noindent To the first order in $\hat{t}_{ji}/U$, this tail is given by
\noindent
\begin{equation}
| \alpha_{i \to j}^{o} \rangle = -\frac{1}{U} | \alpha_{j}^{u} \rangle \langle \alpha_{j}^{u} | \hat{t}_{ji} | \alpha_{i}^{o} \rangle,
\label{eq:wtails}
\end{equation}
\noindent where $| \alpha_{j}^{u} \rangle$ is the unoccupied state in the atomic limit. Then, using Eqs.~(\ref{eq:wfocc}) and (\ref{eq:wtails}), it is straightforward to evaluate the energy change caused by $\hat{t}_{ij}$. For the dimer in Fig.~\ref{fig:ireason}, it is given by
\noindent
\begin{equation}
{\cal E} = -\frac{2 |\langle \alpha_{1}^{o} | \hat{t}_{12} | \alpha_{2}^{u} \rangle|^2}{U}  ,
\label{eq:epert}
\end{equation}
\noindent which is totally consistent with the well-known expression in the SE theory~\cite{Anderson1959}. Nevertheless, the advantage of using the perturbation theory for the Wannier function is that it can be directly applied to the polarization (\ref{eq:elpol}), which can be now expressed via the combinations of $\langle \alpha_{1}^{o} | \hat{t}_{12} | \alpha_{2}^{u} \rangle$ with similar matrix elements of $\vec{r}$ between occupied and unoccupied states at the sites $1$ and $2$~\cite{PRL2021}:
\noindent
\begin{widetext}
\begin{eqnarray}
\vec{P} & = & \frac{e}{VU} \left\{ \langle \alpha_{1}^{o} | \hat{\vec{r}}_{12} | \alpha_{2}^{u} \rangle \langle \alpha_{2}^{u} | \hat{t}_{21} | \alpha_{1}^{o} \rangle + \langle \alpha_{1}^{o} | \hat{t}_{12} | \alpha_{2}^{u} \rangle \langle \alpha_{2}^{u} | \hat{\vec{r}}_{21} | \alpha_{1}^{o} \rangle \right. \nonumber\\
 & & \left. {} + \langle \alpha_{2}^{o} | \hat{\vec{r}}_{21} | \alpha_{1}^{u} \rangle \langle \alpha_{1}^{u} | \hat{t}_{12} | \alpha_{2}^{o} \rangle + \langle \alpha_{2}^{o} | \hat{t}_{21} | \alpha_{1}^{u} \rangle \langle \alpha_{1}^{u} | \hat{\vec{r}}_{12} | \alpha_{2}^{o} \rangle \right\}. \label{eq:pasrt}
\end{eqnarray}
\end{widetext}

\par Then, we need to specify the $o$ and $u$ states and relate them to the direction of spin (or pseudospin). Suppose, there is some basis of Kramers states, $| + \rangle$ and $| - \rangle$, so that $\hat{t}_{12}$ and $\hat{\vec{r}}_{12}$ in Eq.~(\ref{eq:pasrt}) are the matrices in the basis of these states. Then, one can choose
\noindent
\begin{equation}
| \alpha_{i}^{o} \rangle = \cos \frac{\theta_{i}}{2} | i + \rangle + \sin \frac{\theta_{i}}{2} \, e^{ i \phi_{i}}  | i - \rangle ,
\label{eq:o}
\end{equation}
\noindent such that $ \langle \alpha_{i}^{o} | \hat{\boldsymbol{\sigma}} | \alpha_{i}^{o} \rangle = (\sin \theta_{i} \cos \phi_{i}, \sin \theta_{i} \sin \phi_{i}, \cos \theta_{i}) \equiv \boldsymbol{e}_{i}$ is the direction of spin at the site $i$ and $\hat{\boldsymbol{\sigma}} = (\hat{\sigma}^{x},\hat{\sigma}^{y},\hat{\sigma}^{z})$ is the vector of Pauli matrices, where $| + \rangle$ and $| - \rangle$ are the eigenstates of $\hat{\sigma}_{z}$: $\hat{\sigma}_{z} | + \rangle = | + \rangle$ and $\hat{\sigma}_{z} | - \rangle = - | - \rangle$. The $u$ state can be chosen as
\noindent
\begin{equation}
| \alpha_{i}^{u} \rangle =- \sin \frac{\theta_{i}}{2} \, e^{-i \phi_{i}}  | i + \rangle + \cos \frac{\theta_{i}}{2}  | i - \rangle.
\label{eq:u}
\end{equation}

\par Then, $\hat{t}_{12}$ and $\hat{\vec{r}}_{12}$ in Eq.~(\ref{eq:pasrt}) can be decomposed in terms of the $2$$\times$$2$ unity matrix $\hat{\mathbb{1}}$ and the vector $\hat{\boldsymbol{\sigma}}$ of Pauli matrices as $\hat{t}_{12} = t_{12}^{0} \hat{\mathbb{1}} + i \boldsymbol{t}_{12}^{\phantom{0}} \hat{\boldsymbol{\sigma}}$ and $\hat{\vec{r}}_{12} = \vec{r}_{12}^{\, 0} \hat{\mathbb{1}} + i \vec{\boldsymbol{r}}_{12}^{\phantom{0}} \hat{\boldsymbol{\sigma}}$ with the real coefficients $(t_{12}^{0},\boldsymbol{t}_{12}^{\phantom{0}})$ and $(\vec{r}_{12}^{\, 0},\vec{\boldsymbol{r}}_{12}^{\phantom{0}})$. Strictly speaking, the consideration of a single Kramers doublet corresponds to the spin $S = \frac{1}{2}$. Nevertheless, the main conclusions are believed to be more general~\cite{KNB}.

\par Substituting $\hat{t}_{12}$ and $\hat{\vec{r}}_{12}$ in Eq.~(\ref{eq:pasrt}) and using the explicit form of the $o$ and $u$ states, given by Eqs.~(\ref{eq:o})-(\ref{eq:u}), the magnetic part of the exchange energy and polarization can be presented in the form of bilinear interactions between $\boldsymbol{e}_{1}$ and $\boldsymbol{e}_{2}$, which can be further decomposed into isotropic, antisymmetric, and traceless symmetric anisotropic parts as~\cite{PRL2021}
\noindent
\begin{equation}
\mathcal{E} = - J_{12}\boldsymbol{e}_{1} \cdot \boldsymbol{e}_{2} + \boldsymbol{D}_{12} \cdot [\boldsymbol{e}_{1}\times\boldsymbol{e}_{2}] + \boldsymbol{e}_{1} \cdot \mathbb{\Gamma}_{12} \boldsymbol{e}_{2}
\label{eq:spine}
\end{equation}
\noindent and
\noindent
\begin{equation}
\vec{P} = \vec{\mathsf{P}}_{12} \, \boldsymbol{e}_{1} \cdot \boldsymbol{e}_{2} + \vec{\boldsymbol{\mathcal{P}}}_{12} \cdot [ \boldsymbol{e}_{1} \times \boldsymbol{e}_{2} ] + \boldsymbol{e}_{1} \cdot \vec{\mathbb{\Pi}}_{12} \boldsymbol{e}_{2} .
\label{eq:spinp}
\end{equation}
\noindent Here, we use the following conventions: the bold roman letters, $\boldsymbol{D}_{12}$ and $\vec{\boldsymbol{\mathcal{P}}}_{12}$, stand for the vectors forming the scalar product with the spin vector $[ \boldsymbol{e}_{1} \times \boldsymbol{e}_{2} ]$; the hollow symbols, $\mathbb{\Gamma}_{12}$ and $\vec{\mathbb{\Pi}}_{12}$, stand for the $3$$\times$$3$ tensors, acting on $\boldsymbol{e}_{1}$ from the left and $\boldsymbol{e}_{2}$ from the right. The additional vector sign in the case of $\vec{\mathsf{P}}_{12}$, $\vec{\boldsymbol{\mathcal{P}}}_{12}$, and $\vec{\mathbb{\Pi}}_{12}$ denotes the direction of polarization. 

\par The corresponding parameters are summarized in Table~\ref{tab:1orb}~\cite{PRL2021}.
\noindent
\begin{table*}[t]
\caption{Isotropic ($J_{12}$ and $\vec{\mathsf{P}}_{12}$), antisymmetric ($\boldsymbol{D}_{12}$ and $\vec{\boldsymbol{\mathcal{P}}}_{12}$), and traceless anisotropic symmetric ($\mathbb{\Gamma}_{12}$ and $\vec{\mathbb{\Pi}}_{12}$) parameters of exchange interactions and electric polarization. ${\rm Tr}$ denotes trace and $\boldsymbol{\mathds{1}}$ is the $3$$\times$$3$ unity matrix.}
\label{tab:1orb}
\begin{ruledtabular}
\begin{tabular}{ll}
       exchange interactions          & electric polarization \\
\hline \noalign{\vskip 2mm}
$\displaystyle J_{12} = -\frac{(t_{12}^{0})^2}{U}$    & $\displaystyle \vec{\mathsf{P}}_{12} = -\frac{2e}{V} \frac{\vec{r}_{12}^{\, 0} t_{12}^{0}}{U}$         \\ \noalign{\vskip 2mm}
$\displaystyle \boldsymbol{D}_{12} = \frac{2t_{12}^{0}\boldsymbol{t}_{12}^{\phantom{0}}}{U}$    & $\displaystyle \vec{\boldsymbol{\mathcal{P}}}_{12} = -\frac{2e}{V} \frac{ \vec{r}_{12}^{\, 0} \boldsymbol{t}_{12}^{\phantom{0}} + t_{12}^{0} \vec{\boldsymbol{r}}_{12}^{\phantom{0}}}{U}$         \\ \noalign{\vskip 2mm}
$\displaystyle \mathbb{\Gamma}_{12} = \frac{2 \boldsymbol{t}_{12} \otimes \boldsymbol{t}_{12} - {\rm Tr} \left\{ \boldsymbol{t}_{12} \otimes \boldsymbol{t}_{12} \right\} \boldsymbol{\mathds{1}}}{U}$ & $\displaystyle \vec{\mathbb{\Pi}}_{12} = -\frac{e}{V} \frac{ \vec{\boldsymbol{r}}_{12} \otimes \boldsymbol{t}_{12} +  \boldsymbol{t}_{12} \otimes \vec{\boldsymbol{r}}_{12} - {\rm Tr} \{ \vec{\boldsymbol{r}}_{12} \otimes \boldsymbol{t}_{12} \} \boldsymbol{\mathds{1}} }{U}$
\end{tabular}
\end{ruledtabular}
\end{table*}
\noindent These are very general properties, which hold irrespectively on whether there is the inversion center connecting two atomic sites or not.

\par The inversion symmetry imposes an additional constraint on $\hat{t}_{12}$ and $\hat{\vec{r}}_{12}$. Namely, since $\hat{t}_{12}$ is the (true) scalar and $\hat{\vec{r}}_{12}$ is the vector, the former should remain invariant under the permutation of atomic sites, while the latter changes its sign. Therefore, we should have:
\noindent
\begin{equation}
\begin{array}{lll}
t_{21}^{0} = t_{12}^{0}                    & & \boldsymbol{t}_{21} = \boldsymbol{t}_{12} \\
\vec{r}_{21}^{\, 0} = -\vec{r}_{12}^{\, 0} & & \vec{\boldsymbol{r}}_{21} = -\vec{\boldsymbol{r}}_{12}.
\end{array}
\label{eq:inv}
\end{equation}
\noindent Combining these properties with the hermiticity of $\hat{t}_{12}$ and $\hat{\vec{r}}_{12}$, 
\noindent
\begin{equation}
\begin{array}{lll}
t_{21}^{0} = t_{12}^{0}                   & & \boldsymbol{t}_{21} = -\boldsymbol{t}_{12} \\
\vec{r}_{21}^{\, 0} = \vec{r}_{12}^{\, 0} & & \vec{\boldsymbol{r}}_{21} = -\vec{\boldsymbol{r}}_{12},
\end{array}
\label{eq:herm}
\end{equation}
\noindent we find that, in the centrosymmetric case, $\boldsymbol{t}_{12} = 0$ and $\vec{r}_{12}^{\, 0} = 0$. Thus, amongst all possible contributions to exchange interactions and electric polarization (Table~\ref{tab:1orb}), the finite ones will be only the isotropic exchange 
\noindent
\begin{equation}
J_{12} = -\frac{(t_{12}^{0})^2}{U} 
\label{eq:J12}
\end{equation}
\noindent and the antisymmetric (centrosymmetric) contribution to the polarization driven by the parameter 
\noindent
\begin{equation}
\vec{\boldsymbol{\mathcal{P}}}_{12}^{C} = -\frac{2e}{V} \frac{t_{12}^{0} \vec{\boldsymbol{r}}_{12}^{\phantom{0}}}{U}. 
\label{eq:P12C}
\end{equation}
\noindent Another (noncentrosymmetric) contribution to the antisymmetric coupling, 
\noindent
\begin{equation}
\vec{\boldsymbol{\mathcal{P}}}_{12}^{N} = -\frac{2e}{V} \frac{\vec{r}_{12}^{\, 0} \boldsymbol{t}_{12}^{\phantom{0}}}{U}, 
\label{eq:P12A}
\end{equation}
\noindent will vanish, so as other contributions to the polarization of isotropic and symmetric anisotropic types. $\vec{\boldsymbol{\mathcal{P}}}_{12}^{N}$ is entirely related to the crystallographic breaking of inversion symmetry in the bond and proportional to the DM interaction, $\vec{\boldsymbol{\mathcal{P}}}_{12}^{N} \propto \vec{r}_{12}^{\, 0} \, \boldsymbol{D}_{12}^{\, \phantom{0}}$. 

\par Thus, the noncollinear alignment of spins in the centrosymmetric bond will induce the polarization $\vec{P} =  \vec{\boldsymbol{\mathcal{P}}}_{12}^{C} \cdot [ \boldsymbol{e}_{1} \times \boldsymbol{e}_{2} ]$. If the exchange coupling $J_{12}$ is antiferromagnetic ($<$$0$), the noncollinearity itself can be induced by the magnetic field $\boldsymbol{H}$, which is the essence of the conventional magnetoelectric effect resulting in $\vec{P} \sim \boldsymbol{H}$~\cite{DzyaloshinskiiME}. In periodic solids, the noncollinearity may arise from the competition of isotropic exchange interactions $J_{ij}$ in several bonds. Such situation is realized in multiferroic materials of the spin-spiral type~\cite{KNB}. Nevertheless, in both cases, the microscopic origin of the electric polarization is the same and the polarization itself is proportional to $t_{ij}^{0} \vec{\boldsymbol{r}}_{ij}^{\phantom{0}}$. 

\par Alternatively, the external electric field, $\vec{E}$, interacting with $\vec{\boldsymbol{\mathcal{P}}}_{12}^{C}$, will produce the DM interaction $\boldsymbol{D}_{12} = -\vec{E} \cdot \vec{\boldsymbol{\mathcal{P}}}_{12}^{C}$, resulting in a noncollinear alignment of spins. Thus, the DM interactions and the magnetically induced polarization are closely related to each other. The microscopic mechanism of electric polarization induced by the noncollinear alignment of spins is sometimes called the inverse DM mechanism. 

\par The first microscopic theory of electric polarization induced by the noncollinear alignment of spins has been proposed by KNB~\cite{KNB}, which also deals with the behavior of position operator $\vec{r}$ in the basis of Kramers states. However, the KNB theory was also supplemented with several additional assumptions, which are, in principle, not needed for understanding the basic microscopic mechanism:
\begin{itemize}
\item The oxygen states, which were explicitly considered by KNB~\cite{KNB}, do not play an essential role for understanding the microscopic mechanism of emergence the electric polarization and can be eliminated;
\item KNB have considered a very special type of the Kramers doublet: $j = \frac{1}{2}$ states of the $\Gamma_{7}$ symmetry, which are split off by the spin-orbit interaction in the $t_{2g}$ manifold. Formally, such assumption is not necessary as the behavior of electric polarization is related to more fundamental properties of $\hat{t}_{12}$ and $\hat{\vec{r}}_{12}$ under the spatial inversion. Furthermore, the behavior of electric polarization depends on the symmetry of the Kramers states and appears to be less universal, as it may follow from the KNB paper~\cite{KNB}. We will return to this point in Sec.~\ref{sec:sym}.
\end{itemize}

\section{\label{sec:oo} Lack of crystallographic inversion symmetry \\ as manifestation of antiferro orbital ordering}
\par In this section we will briefly discuss the connection of additional term $\vec{\boldsymbol{\mathcal{P}}}_{12}^{N}$, which appears in the noncentrosymmetric case, with the orbital ordering effects~\cite{KugelKhomskii}. The latter are typically considered in the context of Jahn-Teller distortion. They also control the sign of interatomic exchange interactions, known as Goodenough-Kanamori-Anderson (GKA) rules~\cite{Anderson1950,Goodenough1955,Goodenough1958,Kanamori1959}. The spatial inversion about the midpoint of the bond guarantees that the Kramers doublets at two edges of the bond are the same and transformed to each other by $\mathcal{I}$. Since each doublet can be visualized via the distribution of associated with it distribution of electron density, one can say that in this case we deal with the ferro orbital order: in an analogy with the FM spin order, where two spins are the same, here we deal the same electron densities. The electric polarization in this case is controlled solely by $\vec{\boldsymbol{\mathcal{P}}}_{12}^{C}$. The lack of crystallographic inversion symmetry means that the Kramers doublets are different. Therefore, one can say that we deal with some kind of antiferro orbital order, again in terms of distribution of the electron densities. The new aspect in the context of GKA rules is that the antiferro orbital always order breaks the inversion symmetry of each bond~\cite{PRB2024}. Therefore, the antiferro orbital order, which is always noncentrosymmetric, is responsible not only for the FM coupling between the spins, as it is suggested by canonical GKA rules, but also induces the DM interaction and electric polarization in the bond. In the simplest microscopic model, where there is only one Kramers doublet, such polarization is proportional to $\vec{r}_{12}^{\, 0}$ and the term $\vec{\boldsymbol{\mathcal{P}}}_{12}^{N}$ in the expression for the antisymmetric coupling can be viewed as the manifestation of some antiferro orbital order. 

\section{\label{sec:s12} $S=\frac{1}{2}$ and single-ion anisotropies}
\par The fundamental Kramers theorem states that, for the spin-$\frac{1}{2}$, there should be no single-ion contribution to the magnetic energy~\cite{Yosida}. The same property holds for the magnetic dependence of the electric polarization. The proof is extremely simple: for $i=j$ the hermiticity (\ref{eq:herm}) yields: $\boldsymbol{t}_{ii} = 0$ and $\vec{\boldsymbol{r}}_{ii} = 0$, meaning that $\hat{t}_{ii}$ as well as $\hat{\vec{r}}_{ii}$ are simply proportional to the $2$$\times$$2$ unity matrix $\hat{\mathbb{1}}$. Then, neither $\mathbb{\Gamma}_{ii}$ nor $\vec{\mathbb{\Pi}}_{ii}$ depends on the direction of spin. 

\par For multiferroic materials, the single-ion contribution to the electric polarization is typically attributed to the spin-dependent metal-ligand hybridization~\cite{Arima,Murakawa}. The corresponding term is proportional to $(\hat{\boldsymbol{S}} \cdot \vec{\upsilon})^2 \vec{\upsilon}$, where $\vec{\upsilon}$ is the position of ligand atom relative to the transition-metal one with the spin $\hat{\boldsymbol{S}}$. This contribution vanishes if the transition-metal atom is located in the inversion center (for each $\vec{\upsilon}$, there is a contribution with $-\vec{\upsilon}$). For the spin-$\frac{1}{2}$, $\hat{\boldsymbol{S}} = \frac{1}{2} \hat{\boldsymbol{\sigma}}$ and the properties of the Pauli matrices lead to the identity: $(\hat{\boldsymbol{S}} \cdot \vec{\upsilon})^2 \vec{\upsilon} \sim \hat{\mathbb{1}}$.

\section{\label{sec:Kramers} General form of position operator in the basis of Kramers states}
\par Without the loss of generality, two states of the Kramers doublet can be taken in the form:
\noindent
\begin{eqnarray}
| + \rangle & = & | \psi^{0} \uparrow   \rangle + | \Delta \psi \downarrow   \rangle  \label{eq:kramers1} \\
| - \rangle & = & | \psi^{0} \downarrow \rangle - | \Delta \psi^{*} \uparrow \rangle, \label{eq:kramers2} 
\end{eqnarray}
\noindent where the symbols $\uparrow$ and $\downarrow$ explicitly stand for the spin part of these states, while $\psi^{0}$ and $\Delta \psi$ are the coordinate parts. Furthermore, $\psi^{0}$ can be viewed as the main component without SO coupling, while $\Delta \psi$ is induced by the SO coupling. By appropriately choosing the phases, $\psi^{0}$ can be taken as real, while $\Delta \psi$ can be complex. Thus, Eqs.~(\ref{eq:kramers1}) and (\ref{eq:kramers2}) can be formally rewritten as
\noindent
\begin{eqnarray}
| + \rangle & = & | \psi^{0} \uparrow   \rangle + | {\rm Re} \Delta \psi \downarrow \rangle  + i | {\rm Im} \Delta \psi \downarrow \rangle \label{eq:kramers3} \\
| - \rangle & = & | \psi^{0} \downarrow \rangle - | {\rm Re} \Delta \psi \uparrow   \rangle  + i | {\rm Im} \Delta \psi \uparrow   \rangle \label{eq:kramers4} .
\end{eqnarray}
\noindent
If $\mathcal{T} = i \hat{\sigma}^{y} \hat{K}$ is the time-reversal operator (where $K$ stands for the complex conjugation), it holds $\mathcal{T} | + \rangle = - | - \rangle $ and $\mathcal{T} | - \rangle = | + \rangle $. The typical examples are the $j=1/2$ ($\Gamma_{7}$) states~\cite{KNB},
\noindent
\begin{eqnarray}
| + \rangle & = & \frac{1}{\sqrt{3}} \left( | xy \uparrow \rangle + | yz \downarrow \rangle + i | zx \downarrow \rangle \right) \label{eq:jhalfp} \\
| - \rangle & = & \frac{1}{\sqrt{3}} \left( | xy \downarrow \rangle - | yz \uparrow \rangle + i | zx \uparrow \rangle \right) , \label{eq:jhalfm}
\end{eqnarray}
\noindent or the $z^2$ states~\cite{PRL2021}
\noindent
\begin{eqnarray}
| + \rangle & = & \frac{1}{\sqrt{1+2\xi^{2}}} \left( | z^{2} \uparrow \rangle + i\xi | yz \downarrow \rangle +\xi | zx \downarrow \rangle \right) \label{eq:z2p} \\
| - \rangle & = & \frac{1}{\sqrt{1+2\xi^{2}}} \left( | z^{2} \downarrow \rangle + i\xi | yz \uparrow \rangle -\xi | zx \uparrow \rangle \right) \label{eq:z2m}
\end{eqnarray}
\noindent ($\xi$ being the ratio of SO coupling to the crystal field splitting).

\par In order to find the expressions for $\vec{r}_{ij}^{\, 0}$ and $\vec{\boldsymbol{r}}_{ij} = (\vec{r}_{ij}^{\, x},\vec{r}_{ij}^{\, y}, \vec{r}_{ij}^{\, z})$, one have to consider matrix elements of $\vec{r}$ in the basis of the Kramers states $| + \rangle$ and $| - \rangle$ at the sites $i$ and $j$~\cite{SM}:
\noindent
\begin{widetext}
\begin{eqnarray}
\langle i + | \vec{r} | j + \rangle & = & \phantom{-} \langle \psi^{0}_{i} | \vec{r} | \psi^{0}_{j} \rangle + \langle {\rm Re} \Delta \psi_{i}^{\phantom{0}} | \vec{r} | {\rm Re} \Delta  \psi_{j}^{\phantom{0}} \rangle + \langle {\rm Im} \Delta \psi_{i}^{\phantom{0}} | \vec{r} | {\rm Im} \Delta  \psi_{j}^{\phantom{0}} \rangle \nonumber \\
                                    &   & + i \langle {\rm Re} \Delta \psi_{i}^{\phantom{0}} | \vec{r} | {\rm Im} \Delta  \psi_{j}^{\phantom{0}} \rangle -i \langle {\rm Im} \Delta \psi_{i}^{\phantom{0}} | \vec{r} | {\rm Re} \Delta  \psi_{j}^{\phantom{0}} \rangle \label{eq:rijpp} \\
\langle i + | \vec{r} | j - \rangle & = & - \langle \psi^{0}_{i} | \vec{r} | {\rm Re} \Delta  \psi_{j}^{\phantom{0}} \rangle + \langle {\rm Re} \Delta  \psi_{i}^{\phantom{0}} | \vec{r} | \psi^{0}_{j} \rangle + i\langle \psi^{0}_{i} | \vec{r} | {\rm Im} \Delta  \psi_{j}^{\phantom{0}} \rangle -i \langle {\rm Im} \Delta  \psi_{i}^{\phantom{0}} | \vec{r} | \psi^{0}_{j} \rangle \label{eq:rijpm} \\
\langle i - | \vec{r} | j + \rangle & = & \phantom{-} \langle \psi^{0}_{i} | \vec{r} | {\rm Re} \Delta  \psi_{j}^{\phantom{0}} \rangle - \langle {\rm Re} \Delta  \psi_{i}^{\phantom{0}} | \vec{r} | \psi^{0}_{j} \rangle + i\langle \psi^{0}_{i} | \vec{r} | {\rm Im} \Delta  \psi_{j}^{\phantom{0}} \rangle -i \langle {\rm Im} \Delta  \psi_{i}^{\phantom{0}} | \vec{r} | \psi^{0}_{j} \rangle \label{eq:rijmp} \\
\langle i - | \vec{r} | j - \rangle & = & \phantom{-} \langle \psi^{0}_{i} | \vec{r} | \psi^{0}_{j} \rangle + \langle {\rm Re} \Delta \psi_{i}^{\phantom{0}} | \vec{r} | {\rm Re} \Delta  \psi_{j}^{\phantom{0}} \rangle + \langle {\rm Im} \Delta \psi_{i}^{\phantom{0}} | \vec{r} | {\rm Im} \Delta  \psi_{j}^{\phantom{0}} \rangle \nonumber \\
                                    &   & - i \langle {\rm Re} \Delta \psi_{i}^{\phantom{0}} | \vec{r} | {\rm Im} \Delta  \psi_{j}^{\phantom{0}} \rangle + i \langle {\rm Im} \Delta \psi_{i}^{\phantom{0}} | \vec{r} | {\rm Re} \Delta  \psi_{j}^{\phantom{0}} \rangle . \label{eq:rijmm}
\end{eqnarray}
\noindent Then, decomposing
\begin{equation}
\hat{\vec{r}}_{ij} = 
\left(
\begin{array}{cc}
\langle i + | \vec{r} | j + \rangle & \langle i + | \vec{r} | j - \rangle \\
\langle i - | \vec{r} | j + \rangle & \langle i - | \vec{r} | j - \rangle
\end{array}
\right)
\end{equation}
\noindent in terms of the unity $\hat{\mathbb{1}}$ and the vector of Pauli matrices $\hat{\boldsymbol{\sigma}}$, it is straightforward to find that
\begin{eqnarray}
\vec{r}_{ij}^{\, 0} & = & \phantom{-} \langle \psi^{0}_{i} | \vec{r} | \psi^{0}_{j} \rangle + \langle {\rm Re} \Delta \psi_{i}^{\phantom{0}} | \vec{r} | {\rm Re} \Delta  \psi_{j}^{\phantom{0}} \rangle + \langle {\rm Im} \Delta \psi_{i}^{\phantom{0}} | \vec{r} | {\rm Im} \Delta  \psi_{j}^{\phantom{0}} \rangle , \\
\vec{r}_{ij}^{\, x } & = & \phantom{-} \langle \psi^{0}_{i} | \vec{r} | {\rm Im} \Delta  \psi_{j}^{\phantom{0}} \rangle - \langle {\rm Im} \Delta  \psi_{i}^{\phantom{0}} | \vec{r} | \psi^{0}_{j} \rangle , \label{eq:rijx} \\
\vec{r}_{ij}^{\, y } & = & - \langle \psi^{0}_{i} | \vec{r} | {\rm Re} \Delta  \psi_{j}^{\phantom{0}} \rangle + \langle {\rm Re} \Delta  \psi_{i}^{\phantom{0}} | \vec{r} | \psi^{0}_{j} \rangle , \label{eq:rijy} \\
\vec{r}_{ij}^{\, z } & = & \phantom{-} \langle {\rm Re} \Delta \psi_{i}^{\phantom{0}} | \vec{r} | {\rm Im} \Delta  \psi_{j}^{\phantom{0}} \rangle - \langle {\rm Im} \Delta \psi_{i}^{\phantom{0}} | \vec{r} | {\rm Re} \Delta  \psi_{j}^{\phantom{0}} \rangle  \label{eq:rijz} .
\end{eqnarray}
\end{widetext}
\noindent As expected (see Sec.~\ref{sec:microscopic}), $\vec{r}_{ij}^{\, 0}$ is symmetric with respect to the permutation of $i$ and $j$: $\vec{r}_{ji}^{\, 0} = \vec{r}_{ij}^{\, 0}$, while $\vec{\boldsymbol{r}}_{ij}$ is antisymmetric: $\vec{\boldsymbol{r}}_{ji} = -\vec{\boldsymbol{r}}_{ij}$.

\par If ${\rm Re} \Delta  \psi_{j}^{\phantom{0}}$ and ${\rm Im} \Delta  \psi_{j}^{\phantom{0}}$ are induced by the SO interaction, $\vec{r}_{ij}^{\, x }$ and $\vec{r}_{ij}^{\, y }$ are of the first order in the SO coupling, while $\vec{r}_{ij}^{\, z }$ is of the second order.

\section{\label{sec:sym} Symmetry properties of position operator in the basis of Kramers states}
\par Since the electric polarization in the centrosymmetric bonds is proportional to $\vec{\boldsymbol{r}}_{ij}^{\phantom{0}}$, in this sections we closely consider the symmetry properties of this tensor. The bond $ij$ is assumed to be along the $z$ axis, unless it is specified otherwise. 

\subsection{\label{sec:crsym} Crystallographic symmetry properties}
\par First, let us consider the properties of $\vec{\boldsymbol{r}}_{ij}$ under the symmetry operations of the groups $C_{4}$ and $C_{2}$. $C_{4}$ contains four symmetry elements: the $90^{\circ}$-, $180^{\circ}$-, and $270^{\circ}$-rotations about $z$ (respectively, $\hat{R}^{1}$, $\hat{R}^{2}$, and $\hat{R}^{3}$), and the unity $\hat{R}^{0} = \hat{E}$. $C_{2}$ is just the subgroup of $C_{4}$, containing only $\hat{E}$ and $\hat{R}^{2}$. Our main goal here is to elucidate the origin of the differences in the properties of $\vec{\boldsymbol{r}}_{ij}$ when the symmetry is lowed from $C_{4}$ to $C_{2}$. Understanding these differences is crucial for understanding limitations of the KNB expression $\vec{P} \sim \vec{\epsilon}_{ji} \times [\boldsymbol{e}_{i} \times \boldsymbol{e}_{j}]$ for the magnetically induced polarization, where $\vec{\epsilon}_{ji}$ is the unit vector in the direction of the bond~\cite{KNB}.

\par Each symmetry operation $\hat{R}^{n}$ combines rotations in spin ($S$) and orbital ($L$) subspaces: $\hat{R}^{n} = \hat{R}^{n}_{L} \hat{R}^{n}_{S}$, where $\hat{R}^{n}_{S}$ are the $2$$\times$$2$ matrices of ${\rm SU(2)}$ rotations~\cite{LL}, which are listed in Table~\ref{tab:symc4}.
\noindent
\begin{table}[b]
\caption{Transformations under the $90^{\circ}$-, $180^{\circ}$-, and $270^{\circ}$-rotations about $z$. Each symmetry operation combines rotations in spin ($S$) and orbital ($L$) subspaces: $\hat{R}^{n} = \hat{R}^{n}_{S} \hat{R}^{n}_{L}$. The spin matrices $\hat{R}^{n}_{S}$ are listed in the right column of the Table. They specify how $\hat{R}^{n}_{L}$ act on the coordinate part $({\rm Re} \Delta \psi,{\rm Im} \Delta \psi)$ of Kramers states. The same $\hat{R}^{n}_{L}$ describes the transformation of $x$, $y$, and $z$. The upper sign in $({\rm Re} \Delta \psi,{\rm Im} \Delta \psi)$ corresponds to the even representation of the group $C_{4}$, while the lower sign corresponds to the odd one.}
\label{tab:symc4}
\begin{ruledtabular}
\begin{tabular}{clcc}
rotation      & $\hat{R}^{n}_{L}(x,y,z)$ & $\hat{R}^{n}_{L}({\rm Re} \Delta \psi,{\rm Im} \Delta \psi)$  & $\hat{R}^{n}_{S}$ \\
\hline \noalign{\vskip 2mm} 
$90^{\circ}$ & $(-y,x,z)$ &  $(\mp {\rm Im} \Delta \psi,\pm{\rm Re} \Delta \psi)$  &  
$\left(
\begin{array}{cc}
e^{i\frac{\pi}{4}} & 0 \\
0           & e^{-i\frac{\pi}{4}}
\end{array}
\right)$  \\ \noalign{\vskip 2mm} 
$180^{\circ}$ & $(-x,-y,z)$ &  $(-{\rm Re} \Delta \psi, -{\rm Im} \Delta \psi)$  &  
$\left(
\begin{array}{cc}
e^{i\frac{\pi}{2}} & 0 \\
0           & e^{-i\frac{\pi}{2}} \\
\end{array}
\right)$   \\  \noalign{\vskip 2mm} 
$270^{\circ}$ & $(y,-x,z)$ &  $(\pm{\rm Im} \Delta \psi,\mp{\rm Re} \Delta \psi)$ &  
$\left(
\begin{array}{cc}
e^{ i \frac{3\pi}{4}} & 0 \\
0           & e^{-i \frac{3\pi}{4}}
\end{array}
\right)$ 
\end{tabular}
\end{ruledtabular}
\end{table}
\noindent Then, using this explicit form of $\hat{R}^{n}_{S}$, one can find how the coordinate part of the Kramers states will be transformed by $\hat{R}^{n}_{L}$. Indeed, acting by $\hat{R}^{1}$ on $| \pm \rangle$, one finds: 
\noindent
\begin{eqnarray}
\hat{R}^{1} | + \rangle & = & e^{ i\frac{\pi}{4}} \hat{R}^{1}_{L} \left( | \psi^{0} \uparrow   \rangle - i | \Delta \psi \downarrow   \rangle \right) \nonumber\\
\hat{R}^{1} | - \rangle & = & e^{-i\frac{\pi}{4}} \hat{R}^{1}_{L} \left( | \psi^{0} \downarrow \rangle - i | \Delta \psi^{*} \uparrow   \rangle \right).
\end{eqnarray}
\noindent Moreover, without SO coupling, $\hat{R}^{1}_{L}$ should transform $| \psi^{0} \rangle$ to itself, apart from the phase: $\hat{R}^{1}_{L} | \psi^{0} \rangle = \pm | \psi^{0} \rangle $, where the sign depends on whether $| \psi^{0} \rangle$ forms even or odd representation of the group $C_{4}$ (for instance, it is ``$-$'' for the $j=1/2$ states and ``$+$'' for the $z^2$ ones). Then, the requirement that $\hat{R}^{1} | + \rangle$ should coincide with $| + \rangle$ apart from the phase, leads to the following symmetry properties for the coordinate part:
\noindent
\begin{eqnarray}
\hat{R}^{1}_{L} | {\rm Re} \Delta \psi \rangle & = & \mp | {\rm Im} \Delta \psi \rangle \nonumber\\
\hat{R}^{1}_{L} | {\rm Im} \Delta \psi \rangle & = & \pm | {\rm Re} \Delta \psi \rangle.
\end{eqnarray}
\noindent The properties for $\hat{R}^{2}$ and $\hat{R}^{3}$ can be obtained in the same way. The details can be found in Supplemental Material~\cite{SM} and the results are summarized in Table~\ref{tab:symc4} together with the transformation of $x$, $y$, and $z$ by $\hat{R}^{n}_{L}$.

\par Then, we are ready to find which matrix elements of $\vec{\boldsymbol{r}}_{ij}$ remain invariant under the symmetry operations of the group $C_{2}$. Technically, this can be done by symmetrizing $\vec{\boldsymbol{r}}_{ij}$:
\noindent
\begin{equation}
\vec{\boldsymbol{r}}_{ij} \to \frac{1}{2} \sum_{n = 0, 2} \hat{R}^{n}_{L} \vec{\boldsymbol{r}}_{ij},
\end{equation}
\noindent where $\hat{R}^{n}_{L}$ acts both on the coordinate part of the Kramers states and $\vec{r}$. Then, $x$ and $y$ change the sign under $\hat{R}^{2}_{L}$ (see Table~\ref{tab:symc4}), while $z$ does not. Furthermore, the combination of $\psi^{0}$ and either ${\rm Re}$ or ${\rm Im}$ parts of $\Delta \psi$ changes the sign, while any combination of the ${\rm Re}$ or ${\rm Im}$ parts of $\Delta \psi$ does not change the sign. Therefore, the tensor $\vec{\boldsymbol{r}}_{ij}$ should have the following form:
\noindent
\begin{equation}
\vec{\boldsymbol{r}}_{ij} = 
\left(
\begin{array}{ccc}
r_{ij}^{x,x} & r_{ij}^{x,y} & 0 \\
r_{ij}^{y,x} & r_{ij}^{y,y} & 0 \\
0            & 0            & r_{ij}^{z,z}
\end{array}
\right),
\label{eq:rijsymC2}
\end{equation} 
\noindent in agreement with previous analysis~\cite{Xiang,KaplanMahanti}. Thus, $r_{ij}^{z,z}$ can be finite and induce the polarization even if the spins are perpendicular to the bond, so that $[\boldsymbol{e}_{i} \times \boldsymbol{e}_{j}]z$ is parallel to $z$, as in the proper screw spiral. The polarization in this case is also expected to be parallel to $z$. Nevertheless, the effect is expected to be small, being only of the second order in the SO coupling (see Sec.~\ref{sec:Kramers}). The $xy$ block of $\vec{\boldsymbol{r}}_{ij}$ for the $C_{2}$ symmetry does not obey any symmetry properties.

\par For the $C_{4}$ group, the symmetrization takes the form: 
\noindent
\begin{equation}
\vec{\boldsymbol{r}}_{ij} \to \frac{1}{4} \sum_{n = 0}^{3} \hat{R}^{n}_{L} \vec{\boldsymbol{r}}_{ij},
\end{equation}
\noindent where we have to consider two additional symmetry elements $\hat{R}^{1}_{L}$ and $\hat{R}^{3}_{L}$. Then, it is straightforward to find that
\noindent
\begin{equation}
\vec{\boldsymbol{r}}_{ij} = 
\left(
\begin{array}{ccc}
\phantom{-}r_{ij}^{x,x} & -r_{ij}^{y,x} & 0 \\
\phantom{-}r_{ij}^{y,x} & \phantom{-}r_{ij}^{x,x} & 0 \\
0            & 0            & r_{ij}^{z,z}
\end{array}
\right),
\label{eq:rijsymC4}
\end{equation}
\noindent where 
\noindent
\begin{widetext}
\begin{eqnarray}
r_{ij}^{x,x} & = & \frac{1}{2} \left( \langle \psi^{0}_{i} |x| {\rm Im} \Delta \psi_{j} \rangle - \langle {\rm Im} \Delta \psi_{i} |x| \psi^{0}_{j} \rangle - \langle \psi^{0}_{i} |y| {\rm Re} \Delta \psi_{j} \rangle + \langle {\rm Re} \Delta \psi_{i} |y| \psi^{0}_{j} \rangle \right), \\
r_{ij}^{y,x} & = & \frac{1}{2} \left( \langle \psi^{0}_{i} |y| {\rm Im} \Delta \psi_{j} \rangle - \langle {\rm Im} \Delta \psi_{i} |y| \psi^{0}_{j} \rangle + \langle \psi^{0}_{i} |x| {\rm Re} \Delta \psi_{j} \rangle - \langle {\rm Re} \Delta \psi_{i} |x| \psi^{0}_{j} \rangle \right),
\end{eqnarray}
\noindent and
\noindent
\begin{eqnarray}
r_{ij}^{z,z} & = & \phantom{-} \langle {\rm Re} \Delta \psi_{i}^{\phantom{0}} | z | {\rm Im} \Delta  \psi_{j}^{\phantom{0}} \rangle - \langle {\rm Im} \Delta \psi_{i}^{\phantom{0}} | z | {\rm Re} \Delta  \psi_{j}^{\phantom{0}} \rangle .
\end{eqnarray}
\end{widetext}
\noindent Thus, in comparison with the $C_{2}$ group, we have two additional symmetry constraints: $r_{ij}^{x,x} = r_{ij}^{y,y}$ and $r_{ij}^{y,x} = - r_{ij}^{x,y}$. Similar analysis performed for the group $C_{3}$, which includes the $120^{\circ}$ and $240^{\circ}$ rotations about $z$, and $\hat{E}$, yields the same expression for $\vec{\boldsymbol{r}}_{ij}$.

\par Finally, applying symmetry operations of the groups $C_{4}$ and $C_{3}$ to ${\rm Re} \Delta \psi$ and ${\rm Im} \Delta \psi$, it is straightforward to find that the ${\rm Re}$ and ${\rm Im}$ parts of $\Delta \psi$ are orthogonal to each other: 
\noindent
\begin{equation}
\langle {\rm Re} \Delta \psi | {\rm Im} \Delta \psi \rangle = \langle {\rm Im} \Delta \psi | {\rm Re} \Delta \psi \rangle = 0. 
\label{eq:orth}
\end{equation}
\noindent Nevertheless, this is not necessarily the case for the $C_{2}$ symmetry. 

\par These are the properties of $\vec{\boldsymbol{r}}_{ij}$, which are expected from the viewpoint of crystallographic symmetry. Particularly, even for the high symmetries, $C_{3}$ and $C_{4}$, a finite $r_{ij}^{z,z}$ is expected from the crystallographic point of view. This can be noticed in the pioneering work of Dzyaloshinskii~\cite{DzyaloshinskiiME}, who besides the perpendicular magnetoelectric effect in Cr$_2$O$_3$ also predicted the longitudinal one, in the direction of the threefold rotation axis. Nevertheless, besides crystallographic symmetries there are also hidden intrinsic ones, which enforce $r_{ij}^{z,z}$ (and some other matrix elements) to be zero for sufficiently high symmetries in the bond. These effects we will consider in the next section.

\subsection{\label{sec:insym} Hidden intrinsic symmetry properties}
\par This set of properties is related to the explicit dependence of ${\rm Re} \Delta \psi^{\phantom{0}}$ and ${\rm Im} \Delta \psi^{\phantom{0}}$ on the azimuthal angle $\varphi$ in the bond. In the cylindrical coordinates, $x \sim \cos \varphi$ and $y \sim \sin \varphi$, and the $90^{\circ}$-, $180^{\circ}$-, and $270^{\circ}$-rotations about $z$, correspond to the transformations $\varphi \to \varphi + \frac{\pi}{2}$, $\varphi \to \varphi + \pi$, and $\varphi \to \varphi + \frac{3\pi}{2}$, respectively. Then, for the even representation of $C_{4}$, ${\rm Re} \Delta \psi$ and ${\rm Im} \Delta \psi$ transform as, respectively, $x$ and $y$ (see Table~\ref{tab:symc4}). Therefore, it is reasonable to assume that ${\rm Re} \Delta \psi$ and ${\rm Im} \Delta \psi$ should depend on $\varphi$ as $\cos \varphi$ and $\sin \varphi$, respectively, while $\psi^{0}$ does not depend on $\varphi$. For the odd representation, ${\rm Re} \Delta \psi$ and ${\rm Im} \Delta \psi$ transform as, respectively, $y$ and $x$. Therefore, it is reasonable to assume that ${\rm Re} \Delta \psi \sim \sin \varphi$ and ${\rm Im} \Delta \psi \sim \cos \varphi$. Moreover, $\psi^{0}$ for the odd case transform as $\sin 2 \varphi$. Then, the integration over $\varphi$ results in the following property: $r_{ij}^{x,x} = r_{ij}^{y,y} = r_{ij}^{z,z} = 0$, which is valid in the both even and odd case. Therefore, the only nonvanishing element of $\vec{\boldsymbol{r}}_{ij}$ is $r_{ij}^{y,x}$, which is given by
\noindent
\begin{widetext}
\begin{equation}
r_{ij}^{y,x} = \langle \psi^{0}_{i} |y| {\rm Im} \Delta \psi_{j} \rangle - \langle {\rm Im} \Delta \psi_{i} |y| \psi^{0}_{j} \rangle = \langle \psi^{0}_{i} |x| {\rm Re} \Delta \psi_{j} \rangle - \langle {\rm Re} \Delta \psi_{i} |x| \psi^{0}_{j} \rangle . \nonumber
\end{equation}
\end{widetext}

\par Thus, for the $C_{4}$ symmetry, the tensor $\vec{\boldsymbol{r}}_{ij}$ becomes:
\noindent
\begin{equation}
\vec{\boldsymbol{r}}_{ij} = 
\left(
\begin{array}{ccc}
0 & - r_{ij}^{y,x} & 0 \\
\phantom{-} r_{ij}^{y,x} & 0 & 0 \\
0            & 0            & 0
\end{array}
\right),
\label{eq:rijC4}
\end{equation}
\noindent which leads to the well-known expression $\vec{P} \sim \vec{\epsilon}_{ji} \times [\boldsymbol{e}_{i} \times \boldsymbol{e}_{j}]$~\cite{KNB}. The same expression can be obtained for the $C_{3}$ symmetry (see Supplemental Material~\cite{SM} for details). Particularly, since $z$ does not depend on $\varphi$, the property $r_{ij}^{z,z} = 0$ results from the orthogonality condition between ${\rm Re} \Delta \psi$ and ${\rm Im} \Delta \psi$ for the high symmetries $C_{4}$ and $C_{3}$ [see Eq.~(\ref{eq:orth})].

\par For the group $C_{2}$, there is only one symmetry operation, $\hat{R}^{2}_{L}$, which changes the sign of $x$, $y$, ${\rm Re} \Delta \psi$, and ${\rm Im} \Delta \psi$ (see Table~\ref{tab:symc4}). In the other words, all four quantities obey the same transformation law. In such a situation, each of ${\rm Re} \Delta \psi$ and ${\rm Im} \Delta \psi$ can be the linear combination of $\cos \varphi$ and $\sin \varphi$. Then, the tensor $\vec{\boldsymbol{r}}_{ij}$ is generally given by Eq.~(\ref{eq:rijsymC2}) with no additional hidden symmetries, which would reduce the number of its independent matrix elements.

\subsection{\label{sec:order} Order of Kramers states and physical properties}
\par The asymmetric form of the tensor $\vec{\boldsymbol{r}}_{ij}$ is the key result of the KNB theory~\cite{KNB}. Nevertheless, this form depends on the order of the pseudospin states $| + \rangle$ and $| - \rangle$. To certain extend, it is a matter of choice which states are called ``spin up'' and which are ``spin down''. Then, what will happen if one interchanges $| + \rangle$ with $| - \rangle$? Using the expressions for the matrix elements of $\vec{r}$ in the basis of $| + \rangle$ and $| - \rangle$,  Eqs.~(\ref{eq:rijpp})-(\ref{eq:rijmm}), and expanding $\hat{\vec{r}}_{ij}$ in terms of $\hat{\mathbb{1}}$ and $\hat{\boldsymbol{\sigma}}$, it is straightforward to see that the interchange of $| + \rangle$ and $| - \rangle$ will change the sign of $\vec{r}_{ij}^{\, y }$ and $\vec{r}_{ij}^{\, z }$, but not the one of $\vec{r}_{ij}^{\, x }$. Therefore, the tensor $\vec{\tilde{\boldsymbol{r}}}_{ij}$ in this new representation becomes symmetric 
\noindent
\begin{equation}
\vec{\tilde{\boldsymbol{r}}}_{ij} = 
\left(
\begin{array}{ccc}
0 & \tilde{r}_{ij}^{y,x} & 0 \\
\tilde{r}_{ij}^{y,x} & 0 & 0 \\
0            & 0            & 0
\end{array}
\right), 
\end{equation}
\noindent where $\tilde{r}_{ij}^{y,x} = r_{ij}^{y,x}$, but $\tilde{r}_{ij}^{x,y} = -r_{ij}^{x,y} = r_{ij}^{y,x} = \tilde{r}_{ij}^{y,x}$. Such form of the tensor $\vec{\tilde{\boldsymbol{r}}}_{ij}$ was reported in Ref.~\cite{PRL2021} for the $z^{2}$ states. However, this is a consequence of specific order of the states $| + \rangle$ and $| - \rangle$. Therefore, there is no fundamental differences between states $j=\frac{1}{2}$ ($\Gamma_{7}$) and the $z^{2}$: for the given $C_{4}$ symmetry, the tensor $\vec{\boldsymbol{r}}_{ij}$ should obey the same symmetry properties, irrespectively on the type of the basis states. Nevertheless, by changing the order of $| + \rangle$ and $| - \rangle$, one can switch the representations between symmetric and antisymmetric.

\par Such a representation dependence of $\vec{\boldsymbol{r}}_{ij}$ is related to the properties of the pseudospin states. If $|$$\pm$$z$$\rangle \equiv |\pm \rangle$ are the pseudospine states in the positive and negative directions of $z$, corresponding to them pseudospin states in the positive and negative directions of $x$ and $y$ are given by $|$$\pm$$x$$\rangle = \frac{1}{\sqrt{2}}|+ \rangle \pm \frac{1}{\sqrt{2}}|- \rangle$ and $|$$\pm$$y$$\rangle = \pm \frac{1-i}{2}|+ \rangle + \frac{1+i}{2}|- \rangle$, respectively~\cite{PRB2015}. Then, the interchange of $|$$+$$z$$\rangle$ and $|$$-$$z$$\rangle$ will also interchange the states $|$$+$$y$$\rangle$ and $|$$-$$y$$\rangle$ (apart from a phase) but not the states $|$$+$$x$$\rangle$ and $|$$-$$x$$\rangle$, thus explaining why $\vec{r}_{ij}^{\, y }$ and $\vec{r}_{ij}^{\, z }$ change the sign, while $\vec{r}_{ij}^{\, x }$ does not change it. 

\par Of course, the physical properties should not depend on the representation used for the pseudospin states. In reality, the magnetic system is probed by the magnetic field $\boldsymbol{H}$. For instance, such field controls the direction of the spin canting in the conventional magnetoelectric effect~\cite{DzyaloshinskiiME} or the orientation of the spin rotation plane in spiral multiferroics~\cite{TokuraSekiNagaosa}. The interaction $- \mu_{\rm B} \, \boldsymbol{\cal S}_{i} \cdot \mathbb{g}_{i} \boldsymbol{H}$ of the pseudospin $\boldsymbol{\cal S}_{i}$ ($\boldsymbol{\cal S}_{i} = \frac{1}{2}\boldsymbol{e}_{i}$ for the single Kramers doublet) with $\boldsymbol{H}$ is mediated by the $3$$\times$$3$ $g$-tensor $\mathbb{g}_{i}$, which is given by the matrix elements of $\hat{\boldsymbol{\sigma}}$ and the angular momentum operator $\hat{\boldsymbol{L}} = (\hat{L}^{x},\hat{L}^{y},\hat{L}^{z})$ in the basis of pseudospin states~\cite{PRB2015}. Therefore, for the correct description of observable magnetoelectric properties, it is essential to use the same order of the pseudospin states in the definition of \emph{both} tensors $\vec{\boldsymbol{r}}_{ij}$ and $\mathbb{g}_{i}$.

\par
\begin{figure}[b]
\begin{center}
\includegraphics[width=8.6cm]{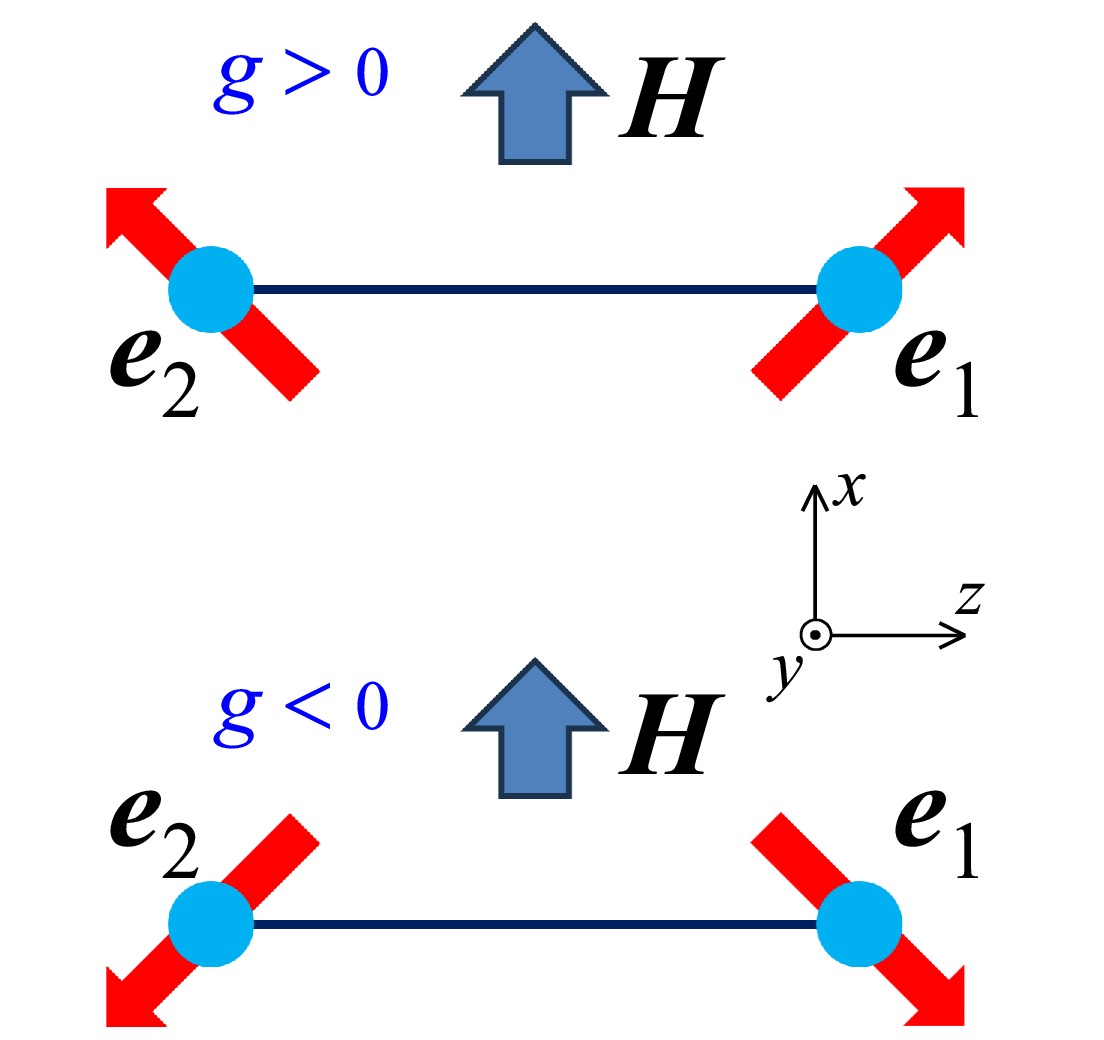}
\end{center}
\caption{A cartoon picture explaining the canting of spins induced by the magnetic field depending on the sign of the g-factor.}
\label{fig:gfactor}
\end{figure}

 As an example, let us consider two canted spins, whose AFM component along $z$ coexists with the FM one in the $xy$ plane (Fig.~\ref{fig:gfactor}). The latter is induced by the magnetic field $\boldsymbol{H}$, which can freely roatte in the $xy$ plane. The corresponding directions of spins are given by $\boldsymbol{e}_{1,2} = (\sin \theta \cos \varphi,\sin \theta \sin \varphi,\pm \cos \theta)$, in terms of the polar ($\theta$) and azimuthal ($\varphi$) angles. 
\noindent
\noindent For $\boldsymbol{H}$ along $y$ ($\varphi = \frac{\pi}{2}$), the tensors $\vec{\boldsymbol{r}}_{12}$ and $\vec{\tilde{\boldsymbol{r}}}_{12}$ produce the same polarization $\vec{P} = (0,P,0)$ along $y$. Then, the rotation of $[\boldsymbol{e}_{1} \times \boldsymbol{e}_{2}] = -\sin 2\theta(\sin \varphi, -\cos \varphi, 0)$ from $x$ ($\varphi = \frac{\pi}{2}$) to $-y$ ($\varphi = 0$) would rotate $\vec{P} = (0,P,0)$ to $\vec{P} = (P,0,0)$ in the case of $\vec{\boldsymbol{r}}_{12}$ and $\vec{P} = (-P,0,0)$ in the case of $\vec{\tilde{\boldsymbol{r}}}_{12}$. The discrepancy is resolved by considering explicitly the g-tensor $\mathbb{g} = [g^{ab}]$ ($a$, $b=$ $x$, $y$, or $z$), which is diagonal and depends on the order of the pseudospin states such that $\tilde{g}^{yy} = g^{yy}$, but $\tilde{g}^{xx} = -g^{xx}$. Then, for $\tilde{\mathbb{g}}$, $\boldsymbol{H} || x$ will additionally change the sign of $[\boldsymbol{e}_{1} \times \boldsymbol{e}_{2}]$ and, together with $\vec{\tilde{\boldsymbol{r}}}_{12}$, produce $\vec{P} = (P,0,0)$, the same as for $\vec{\boldsymbol{r}}_{12}$.

\section{\label{sec:other} Other contributions to $\vec{P}$}
\par In the previous sections have considered the simplest microscopic theory of magnetoelectric coupling where the magnetically active states at each site of the lattice are represented by a single Kramers doublet. The key result of this theory is that, even in the crystallographically centrosymmetric bonds, the noncollinear alignment of spins can break the inversion symmetry and induce the electric polarization. The tensor describing this magnetoelectric coupling, $\vec{\boldsymbol{\mathcal{P}}}_{ij}^{C}$, is given by Eq.~(\ref{eq:P12C}) and its symmetry properties have been considered in Sec.~\ref{sec:sym}. This funding further justifies and generalizes the KNB theory of the magnetoelectric coupling~\cite{KNB}. However, the situation realized in real materials can be more complex and it would not be entirely right to analyse all the cases from the viewpoint of only this simplest model for the magnetoelectric coupling. In fact, there can be other mechanisms and all of them can be rigorously derived starting from the general expression (\ref{eq:elpol}) for the electric polarization in terms of the Wannier functions for the occupied bands~\cite{FE_theory1,FE_theory2,FE_theory3}.

\par First, if the bond is noncentrosymmetric, there will be an additional contribution to $\vec{P}$. The corresponding parameter $\vec{\boldsymbol{\mathcal{P}}}_{ij}^{N}$, describing the antisymmetric magnetoelectric coupling, is given by Eq.~(\ref{eq:P12A}). It is obtained along the same line as $\vec{\boldsymbol{\mathcal{P}}}_{ij}^{C}$, considering the intersite matrix elements of $\vec{r}_{ij}^{\, 0}$ in the basis of Kramers states. $\vec{\boldsymbol{\mathcal{P}}}_{ij}^{N}$ can also operate in the centrosymmetric systems, but where the inversion centers are not necessarily located in the midpoints of the bonds. The typical example is orthorhombic manganites, which will be considered in Sec.~\ref{sec:TbMnO3}. Besides $\vec{\boldsymbol{\mathcal{P}}}_{ij}^{N}$, the isotropic and symmetric anisotropic contributions to the polarization, which are driven by, respectively, $\vec{\mathsf{P}}_{ij}$ and $\vec{\mathbb{\Pi}}_{ij}$, become finite for the noncentrosymmetric bonds.

\par
\begin{figure}[t]
\begin{center}
\includegraphics[width=8.6cm]{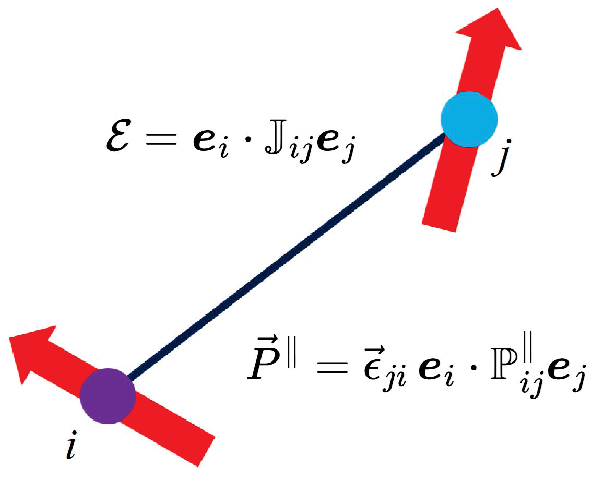}
\end{center}
\caption{Modulation of electric polarization by the magnetic alignment in the noncentrosymmetric dipole. The direction of polarization is fixed along $\vec{\epsilon}_{ji}$ (the unit vector along the dipole). However, the magnitude of the polarization can depend on the directions of spins $\boldsymbol{e}_{i}$ and $\boldsymbol{e}_{j}$ in the dipole, similar to the exchange energy.}
\label{fig:dipole}
\end{figure}

 Even more contributions are expected in the multiorbital case, where there is more than one Kramers doublet associated with each magnetic site. For instance, the noncentrosymmetric bond can be viewed as a dipole. Considering only two magnetic atoms at the edges of this bond ($i$ and $j$ in Fig.~\ref{fig:dipole}) and disregarding explicit contributions of intermediate ligand sites, the electric dipole moment will be parallel to the bond. Mathematically, this approximation is obtained by replacing the $r$-space integration in Eq.~(\ref{eq:elpol}) by the summation over only two points $i$ and $j$. Then, one can control the charge transfer between $i$ and $j$ (and, therefore, the magnitude of the electric dipole moment in the bond) by realigning the spins at the sites $i$ and $j$. 
\noindent
\noindent The exchange energy in the noncentrosymmetric bond is given by $\boldsymbol{e}_{i} \cdot \mathbb{J}_{ij} \boldsymbol{e}_{j}$, where the $3$$\times$$3$ tensor $\mathbb{J}_{ij}$ can be decomposed into the isotropic ($J_{ij}$), DM ($\boldsymbol{D}_{ij}$), and symmetric anisotropic ($\mathbb{\Gamma}_{ij}$) parts. Then, the polarization will have a similar form, $\vec{\epsilon}_{ji} \boldsymbol{e}_{i} \cdot \mathbb{P}_{ij}^{\parallel} \boldsymbol{e}_{j}$, where $\vec{\epsilon}_{ji}$ is the unit vector in the direction of the bond and the $3$$\times$$3$ tensor $\mathbb{P}_{ij}^{\parallel}$ can be further rearranged in terms of the isotropic, antisymmetric (DM-like), and symmetric anisotropic parts~\cite{PRB2019,PRB2020}. 

\par The microscopic mechanism behind $\mathbb{P}_{ij}^{\parallel}$ is essentially multi-orbital one. For each bond, $\mathbb{P}_{ij}^{\parallel}$ is proportional to $J_{\rm H}\left( (t_{ij}^{ab})^2 - (t_{ij}^{ba})^2 \right)$, where $J_{\rm H}$ is the intraatomic exchange (Hund's rule coupling) and $t_{ij}^{ab}$ is the transfer integral between orbitals $a$ and $b$~\cite{PRB2019,PRB2020}. Thus, the diagonal contributions with $a=b$ vanish, so that the considered mechanism does not operate in the one-orbital case. In the multi-orbital case, it is driven by the Hund's rule coupling $J_{\rm H}$.

\section{\label{sec:examples} Examples and classification}
\par In this section we reconsider the classification of some known magnetoelectric and multiferroic materials in terms of main microscopic mechanisms of the magnetoelectric coupling: 
\begin{itemize}
\item The generalized KNB mechanism (gKNB), where the inversion symmetry in the centrocymmetric bond can be broken by a noncollinear magnetic order~\cite{KNB}. Nevertheless, contrary to the conventional KNB mechanism, the direction of electric polarization is not necessary limited by the expression $\vec{P} \sim \vec{\epsilon}_{ji} \times [\boldsymbol{e}_{i} \times \boldsymbol{e}_{j}]$, which can be justified only for relatively high $C_{3}$ or $C_{4}$ symmetries of the bond;
\item The single-ion mechanism, which is a generalization of spin-dependent metal–ligand hybridization mechanism~\cite{Arima,Murakawa};
\item In noncentrosymmetric bonds, other mechanisms can also take place.
\end{itemize}

\par It is true that the KNB mechanism is generic in the sense that it can operate in all types of insulators, which can potentially form a noncollinear magnetic structure. However, there can be many complications caused by other mechanisms and contributions.

\subsection{\label{sec:Cr2O3} Cr$_2$O$_3$}
\par
\begin{figure}[h]
\begin{center}
\includegraphics[width=8.6cm]{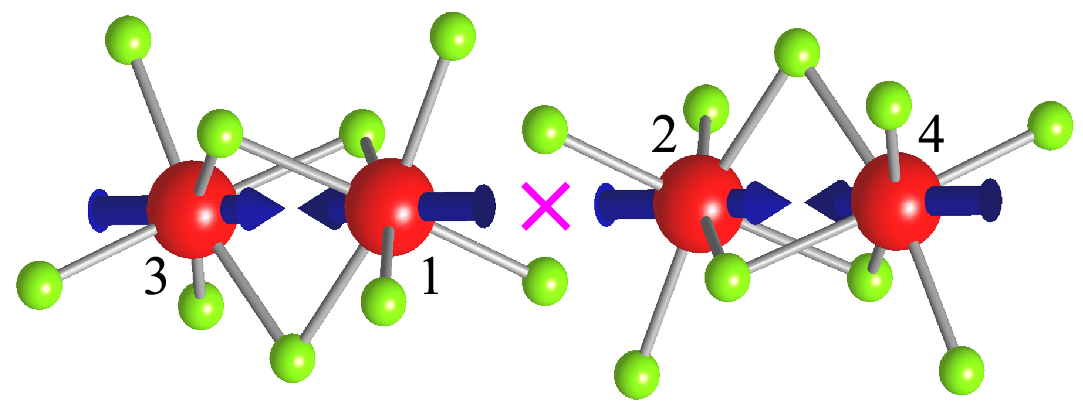}
\end{center}
\caption{
Fragment of crystal and magnetic structure of Cr$_2$O$_3$. The directions of magnetic moments are denoted by arrows. The crystallographic inversion center is denoted by $\times$. Cr and O atoms are denoted by red and green spheres, respectively.}
\label{fig:cr2o3}
\end{figure}

 Cr$_2$O$_3$ is the canonical material exhibiting the magnetoelectric effect, which was theoretically proposed by Dzyaloshinskii in 1959~\cite{DzyaloshinskiiME} and experimentally verified by Astrov in 1960~\cite{Astrov}. Cr$_2$O$_3$ crystallizes in corundum structure (the space group $R\overline{3}c$) with the inversion center connecting two pairs of Cr atoms in the unit cell (see Fig~\ref{fig:cr2o3}). 
\noindent
\noindent The magnetic moments are aligned antiferromagnetically, such that the pairs of sites $1$ and $2$ ($3$ and $4$) are transformed to each other by the symmetry operation $\mathcal{IT}$. This symmetry can be broken by applying the magnetic field perpendicular to the bonds $12$ and $34$, resulting in net electric polarization. The bonds $12$ and $34$ have the $C_{3}$ symmetry. Therefore, the polarization in these bonds is expected to obey the conventional KNB rules, where the tensor $\vec{\boldsymbol{r}}_{ij}$ is given by Eq.~(\ref{eq:rijC4}) and the polarization itself is proportional to $\vec{\epsilon}_{ji} \times [\boldsymbol{e}_{i} \times \boldsymbol{e}_{j}]$~\cite{KNB}. In fact, this is rare case where the conventional KNB expression can be justified. In other multiferroic materials, which will be considered below, the bond symmetry is lower and we generally deal with a more complex form of the tensor $\vec{\boldsymbol{\mathcal{P}}}_{ij}$. In addition to the KNB mechanism, the single-ion contributions are also expected in the case of Cr$_2$O$_3$, as we deal with the $S = \frac{3}{2}$ Cr$^{3+}$ ions located in the noncentrosymmetric positions. Furthermore, the bonds $14$ and $23$ are noncentrosymmetric, that may activate other mechanisms of the magnetoelectric coupling besides KNB and single-ion anisotropy. For instance, the antisymmetric exchange coupling in the bonds $14$ and $23$ is known to be responsible for the phenomenon of weak ferromagnetism in Fe$_2$O$_3$, which has the same crystal structure but different type of the AFM order~\cite{Dzyaloshinskii_weakF}. Therefore, there could be also a finite contribution of these bonds to the electric polarization $\vec{P}$~\cite{PRB2017}. Nevertheless, for the conventional magnetoelectric effect in Cr$_2$O$_3$, the spins in the bonds $14$ and $23$ remain parallel. Therefore, there should be no contributions to $\vec{P}$ arising from these noncentrosymmetric bonds, perhaps except small symmetric anisotropic ones. The results are summarized in Table~\ref{tab:classification}.
\noindent
\begin{table*}[t]
\caption{Summary of main microscopic mechanisms operating (or not) in real magnetoelectric (ME) or multiferroic (MF) materials: the KNB mechanism in centrosymmetric bonds, the single-ion (SI) mechanism, and other mechanisms emerging in noncetrosymmetric bonds. The KNB rules, relating the direction of electric polarization with the cross product of magnetic moments in the bond, are specified in parentheses, where `c' stands for the conventional KNB rules expected for high bond symmetries, and `g' stands for the generalized rules expected for lower symmetries. If the SI mechanism does not operate, the reason is explained in parentheses: $S$$=$$1/2$, or the magnetic ions are located in inversion centers (denoted as `ic'), or both or them.}
\label{tab:classification}
\begin{ruledtabular}
\begin{tabular}{lccccc}
Material            & Space Group           & Type  & KNB     & SI                      & Other \\
\hline
Cr$_2$O$_3$         & $R\overline{3}c$      & ME    & Yes (c) & Yes                     & No    \\
TbMnO$_3$           & $Pbnm$                & MF    & Yes (g) & No (ic)                 & Yes   \\
MnWO$_4$            & $P2/c$                & MF    & Yes (g) & Yes                     & Yes   \\
CuCl$_2$            & $C2/m$                & MF    & Yes (g) & No (ic and $S$$=$$1/2$)    & No    \\
CuO                 & $C2/c$                & MF    & Yes (g) & No ($S$$=$$1/2$)        & Yes   \\
Ba$_2$CoGe$_2$O$_7$ & $P\overline{4}2_{1}m$ & MF    & No      & Yes                     & No    \\
Ba$_2$CuGe$_2$O$_7$ & $P\overline{4}2_{1}m$ & MF    & Yes (g) & No ($S$$=$$1/2$)        & Yes   \\
CuFeO$_2$           & $R\overline{3}m$      & MF    & Yes (g) & No (ic)                 & No    \\
MnI$_2$             & $P\overline{3}m1$     & MF    & Yes (g) & No (ic)                 & No    \\
RbFe(MoO$_4$)$_2$   & $P\overline{3}$       & MF    & Yes (g) & No (ic)                 & No    \\
\end{tabular}
\end{ruledtabular}
\end{table*} 

\subsection{\label{sec:TbMnO3} TbMnO$_3$ and other multiferroic manganites}
\par
\begin{figure}[b]
\begin{center}
\includegraphics[width=8.6cm]{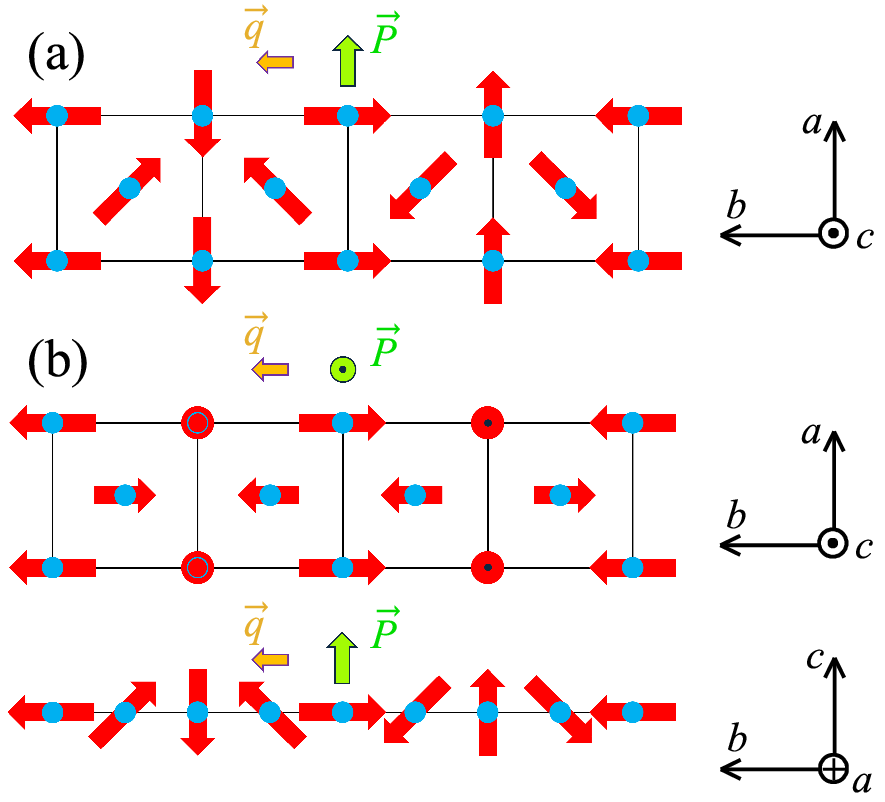}
\end{center}
\caption{
Magnetic cycloidal structures realized in TbMnO$_3$. (a) $ab$ cycloid. (b) $bc$ cycloid: the view in the $ab$ plane (top) and in the $bc$ plane (bottom). $\vec{q}$ shows the propagation direction of the spin spiral. $\vec{P}$ shows the corresponding direction of electric polarization.}
\label{fig:TbMnO3}
\end{figure}

 TbMnO$_3$ is one of the first experimentally discovered multiferroic materials, where cycloidal magnetic order, developed in otherwise centrosymmetric orthorhombic crystal (the space group $Pbnm$), gives rise to spontaneous electric polarization~\cite{Kimura_TbMnO3}. The direction of this polarization depends on the orientation of the cycloidal plane and this dependence is well explained in terms of the conventional KNB theory. Particularly, the spin-spiral propagation vector in TbMnO$_3$ is $\vec{q} \approx (0,\frac{1}{4},0)$. Then, the spin spiral in the orthorhombic $ab$ plane induces the polarization along the $a$ axis, while the spin spiral in the plane $bc$ induces the polarization along the $c$ axis (see Fig.~\ref{fig:TbMnO3}). Thus, in both cases $\vec{P}$ is proportional to $\vec{q} \times [\boldsymbol{e}_{i} \times \boldsymbol{e}_{j}]$, in agreement with the KNB theory~\cite{KNB}.
\noindent

\par In the ideal $Pbnm$ structure, the magnetic Mn$^{3+}$ ions are located in the inversion centers, thus excluding single-ion contributions to the electric polarization. Nevertheless, in all other respects the situation is not such simple: The symmetry of the Mn-O-Mn bonds is low and all these bonds are noncentrosymmetric. Therefore, in addition to the conventional KNB mechanism, there can be other contributions to $\vec{P}$. 

\par
\begin{figure}[b]
\begin{center}
\includegraphics[width=8.6cm]{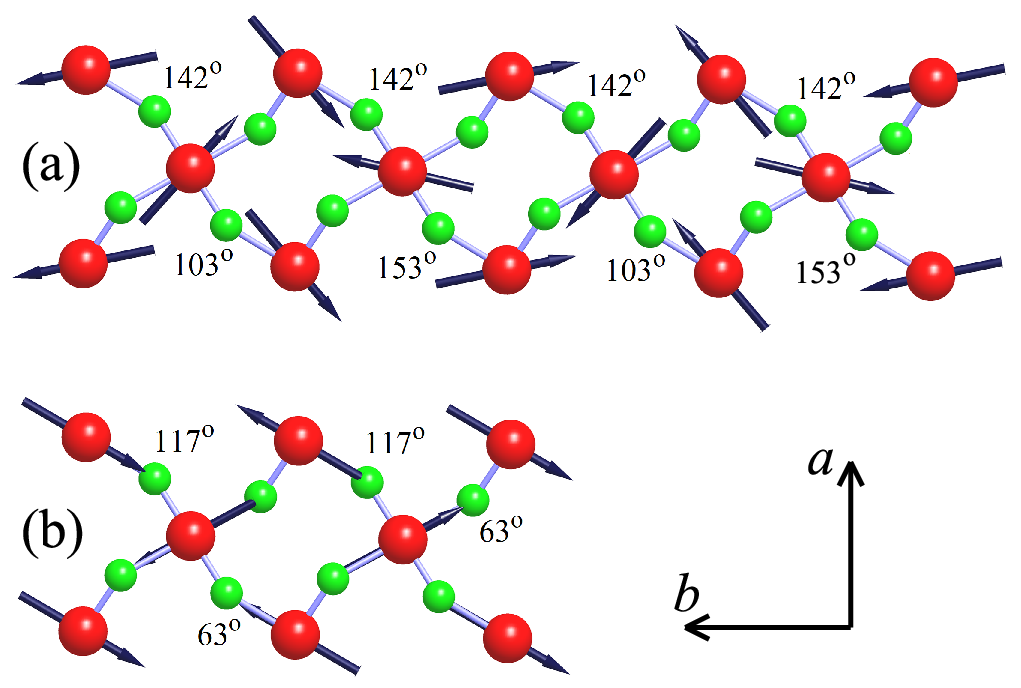}
\end{center}
\caption{
Magnetic textures obtained in the Hartree-Fock approximation for realistic electronic model (constructed for the Mn $3d$ bands in TbMnO$_3$) assuming (a) fourfold magnetic periodicity and (b) twofold magnetic periodicity~\cite{PRB2011}. Mn and O atoms are denoted by red and green spheres, respectively. Mn atoms are located in inversion centers.}
\label{fig:TbMnO3HF}
\end{figure}

 Another complication arises from the fact that the spin-spiral texture in TbMnO$_3$ and related orthorhombic perovskite manganites can be deformed due to the large single-ion anisotropy, which tends to lock the spin spiral to the lattice and make it a commensurate one. If the regular spin-spiral texture is characterized by the same $\boldsymbol{e}_{i} \cdot \boldsymbol{e}_{i}$ in all the bonds, the deformation makes some of them inequivalent, including those, which were crystallographically connected by the spatial inversion (see Fig.~\ref{fig:TbMnO3HF}a)~\cite{PRB2011}. This gives rise to the additional, isotropic, contribution to $\vec{P}$ in the case of the $ab$ cycloid.
\noindent
\noindent This mechanism can be easily rationalized in terms of the double exchange model, which should be applicable for the high-spin ions, such as Mn$^{3+}$ in manganites~\cite{deGennes}. According to this model, the effect of noncollinear alignment of spins in the bond $ij$ can be described by redefining the transfer integral as $\hat{t}_{ij} \to \xi_{ij}\hat{t}_{ij}$, where $|\xi_{ij}| = \sqrt{\frac{1 + \boldsymbol{e}_{i} \cdot \boldsymbol{e}_{j}}{2}}$. Thus, if the bonds on opposite sides of the inversion center are characterized by different $\xi_{ij}$'s, the inversion symmetry will be broken. The mechanism appears to be generic and can be applied to any magnetic texture, which can be viewed as a deformed spin spiral~\cite{PRB2014}. The example of such twofold periodic magnetic texture in orthorhombic manganites is shown in Fig.~\ref{fig:TbMnO3HF}b. It can be viewed as the spin spiral with $\vec{q} = (0,\frac{1}{2},0)$, which is deformed by the single-ion anisotropy term. The latter tends to align the spins along the longest Mn-O bond within the orthorhombic $ab$ plane.

\par
\begin{figure}[b]
\begin{center}
\includegraphics[width=8.6cm]{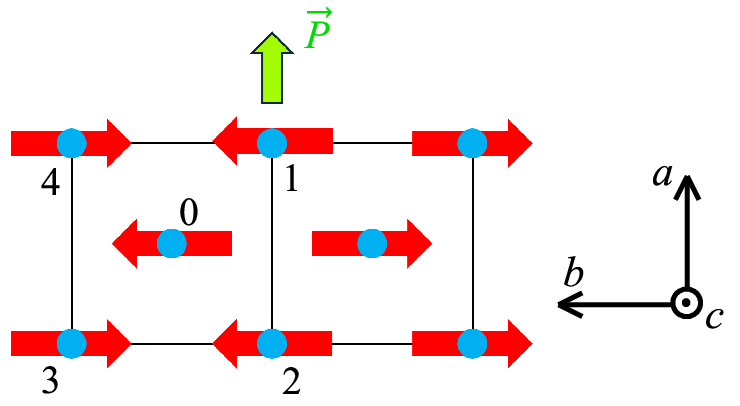}
\end{center}
\caption{
E-type antiferromagnetic texture realized in HoMnO$_3$ and YMnO$_3$. $\vec{P}$ shows the corresponding direction of electric polarization.}
\label{fig:E}
\end{figure}

 The special case of the twofold periodic magnetic texture in manganites is the collinear E-type AFM phase (see Fig.~\ref{fig:E}), which is realized in HoMnO$_3$~\cite{HoMnO3} and YMnO$_3$~\cite{YMnO3}, and can be viewed as a limiting case of the deformed spin spiral with $\vec{q} = (0,\frac{1}{2},0)$.
\noindent
\noindent In this phase, two of the nearest bonds ($01$ and $02$ in Fig.~\ref{fig:E}) are ferromagnetic, while two other bonds ($03$ and $04$, which are connected with $01$ and $02$ by the spatial inversion) are antiferromagnetic. Thus, according to the double exchange mechanism, the inversion symmetry is broken and the system develops a spontaneous polarization. Like in the noncollinear case (see Sec.~\ref{sec:ireason}), the inversion symmetry breaking in the E-phase occurs because some parts of the magnetic pattern are transformed to themselves by the symmetry operation $\mathcal{IT}$, while other parts require $\mathcal{I}$ to be the symmetry operation, which is incompatible with $\mathcal{IT}$. Indeed, the sites $1$ and $3$ ($2$ and $4$) can be transformed to each other by the symmetry operation $\mathcal{IT}$. However, such symmetry implies that the central site $0$ should be nonmagnetic. Although such state is allowed from the symmetry point of view, for the high-spin ions Mn$^{3+}$, it would result in the gigantic loss of the Hund's rule energy. Thus, all Mn sites in the E-type AFM texture remain magnetic, but it breaks the inversion symmetry. 

\par The collinear alignment of spins in the E-phase is enforced by the exchange striction effects, associated with off-centrosymmetric displacements of Mn atoms~\cite{YMnO3,PRB2012}. The lattice displacements (the so-called inverse DM effect) are also expected in the spin-spiral phase of TbMnO$_3$~\cite{SergienkoDagotto,MalashevichVanderbilt}.

\par Thus, although the KNB mechanism is certainly relevant to the physics of TbMnO$_3$ and other multiferroic manganites, there are many other contributions affecting the behavior of electric polarization in this type of compounds. 

\subsection{\label{sec:MnWO4} MnWO$_4$ and related spin-spiral multiferroics with low crystallographic symmetry}
\par There are several spin-spiral multiferroics with relatively low crystallographic symmetry, including Ni$_3$V$_2$O$_8$ (space group $Cmca$)~\cite{Ni3V2O8}, FeVO$_4$ ($P\overline{1}$)~\cite{FeVO4}, and MnWO$_4$ ($P2/c$)~\cite{Taniguchi,Arkenbout,Heyer}. In this section we will briefly discuss some basic features of magnetoelectric coupling in MnWO$_4$, which are inherent to the behavior of other spin-spiral multiferroics with low crystallographic symmetry.  
\noindent
\begin{figure}[t]
\begin{center}
\includegraphics[width=8.6cm]{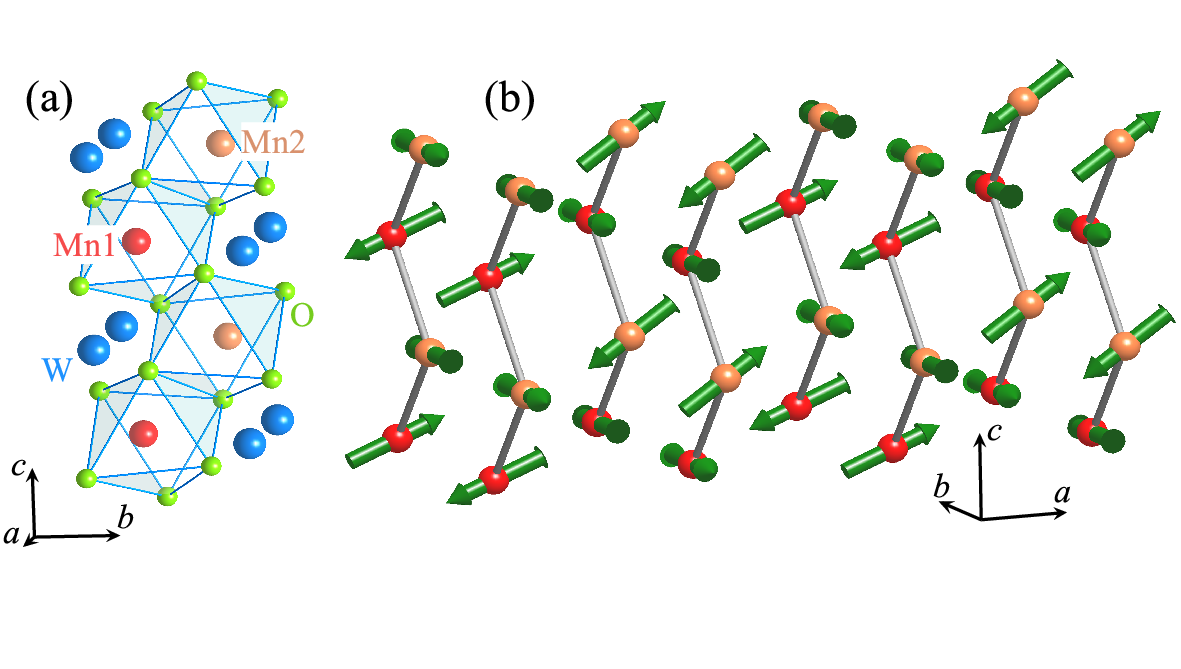}
\end{center}
\caption{
Fragment of crystal (a) and magnetic (b) structure of MnWO$_4$ in the noncollinear phase. Two Mn sublattices are shown by different colors.}
\label{fig:MnWO4}
\end{figure}
\noindent

\par MnWO$_4$ crystallizes in the monoclinic $P2/c$ structure, which is shown in Fig.~\ref{fig:MnWO4}a, and develops the spiral magnetic order in the temperature range $7.6$ K $\lesssim T \lesssim$ $12.5$ K with the propagation vector $\vec{q} \approx (-\frac{1}{4},\frac{1}{2},\frac{1}{2})$, as explained in Fig.~\ref{fig:MnWO4}b. Thus, the spin spiral propagates along the monoclinic $a$ axis, while the magnetic coupling in two other directions is antiferromagnetic. The cross product $\boldsymbol{e}_{i} \times \boldsymbol{e}_{j}$ formed by neighboring spins lie in the $ac$ plane. The spiral magnetic order induces the electric polarization along the $b$ axis, that is formally consistent with the KNB rules~\cite{Taniguchi}. The polarization vanishes in the collinear phases realized below $7.6$ K and above $12.5$ K.  

\par Nevertheless, since the symmetry is low, there can be other contributions to the electric polarization. First, the KNB mechanism itself is expected to be of the generalized from, where other matrix elements of $\vec{\boldsymbol{r}}_{ij}$ can play some role. Then, the $P2/c$ structure of MnWO$_4$ has the inversion center, which connects two Mn sublattices with each other. In Fig.~\ref{fig:MnWO4}, these sublattices are shown by different colors. Therefore, the bonds formed between the sublattices are centrosymmetric, while the bonds within each sublattice are not. In the latter case, there can be other contributions to $\vec{P}$: isotropic, $\vec{\boldsymbol{\mathcal{P}}}_{12}^{N}$, and symmetric anisotropic. 

\par The spin-spiral alignment in MnWO$_4$ is caused by the competition of isotropic exchange interactions~\cite{PRB2013}. The DM interaction is forbidden between the sublattices, but can operate within each sublattice. For each bond $\boldsymbol{R}$ connecting two Mn sites in the sublattice, the vectors of DM interactions in different sublattices are related by the identity $\boldsymbol{D}^{2}(\boldsymbol{R}) = \boldsymbol{D}^{1}(-\boldsymbol{R}) = -\boldsymbol{D}^{1}(\boldsymbol{R})$, which describes the transformation of the axial vector under the spatial inversion. Then, from the viewpoint of DM interactions, the spins in different sublattices prefer to spiral in the opposite directions (say, clockwise and counterclockwise). However, the competing isotropic interactions, which are generally stronger that the DM ones, will establish the same type of spiral in both the sublattices. Thus, in one sublattice, the isotropic and DM contributions to the spin spiral will be collaborative, while in other sublattice they will inevitably be competing. Thus, in the spin-spiral phase, the Mn sublattices become intrinsically nonequivalent, that destroys the spatial inversion and contributes to the behavior of electric polarization.

\subsection{\label{sec:CuCl2} CuCl$_2$ and CuBr$_2$}
\par
\begin{figure}[b]
\begin{center}
\includegraphics[width=5cm]{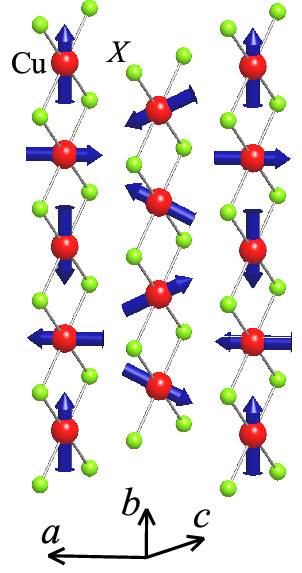}
\end{center}
\caption{
Fragment of crystal and magnetic structure of CuX$_2$ ($X=$ Cl or Br).}
\label{fig:CuX2}
\end{figure}

 Cu$X_2$ ($X=$ Cl or Br) are the only known so far multiferroic materials, where the KNB mechanism operates alone, without additional complications. They crystallize in the monoclinic $C2/m$ structure, consisting of the Cu$X_2$ chains propagating along the $b$ axis~\cite{Seki_CuCl2,CuBr2}. Competing isotropic exchange interactions stabilize the spin-spiral multiferroic phase with $\vec{q} \approx (1,\frac{1}{4},\frac{1}{2})$~\cite{PRL2021,Seki_CuCl2}, as explained in Fig.~\ref{fig:CuX2}. 
\noindent
\noindent
The magnetic Cu$^{2+}$ ions are located in the inversion centers. Moreover, the nominal spin of Cu$^{2+}$ is $S=1/2$. Therefore, there is no single-ion contributions to the electric polarization. All Cu-Cu bonds are centrosymmetric. Nevertheless, the symmetry of the bonds is low: for the bonds in the chains, it is $C_{2}$ (where the rotation axis is $b$), while for other bonds there is no symmetry at all, apart from the inversion. Thus, we deal with the generalized KNB mechanism, where the $3$$\times$$3$ tensors $\vec{\boldsymbol{r}}_{ij}$ is specified by five elements for the bonds parallel to $b$ and all nine elements for other bonds. Nevertheless, the contributions in the chains clearly prevail~\cite{PRL2021}. Then, the polarization is expected to be in the $ac$ plane, depending in the orientation of the spin-spiral plane, which can be controlled by the magnetic field~\cite{Seki_CuCl2}. 

\subsection{\label{sec:CuO} CuO}
\par
\begin{figure}[b]
\begin{center}
\includegraphics[width=8.6cm]{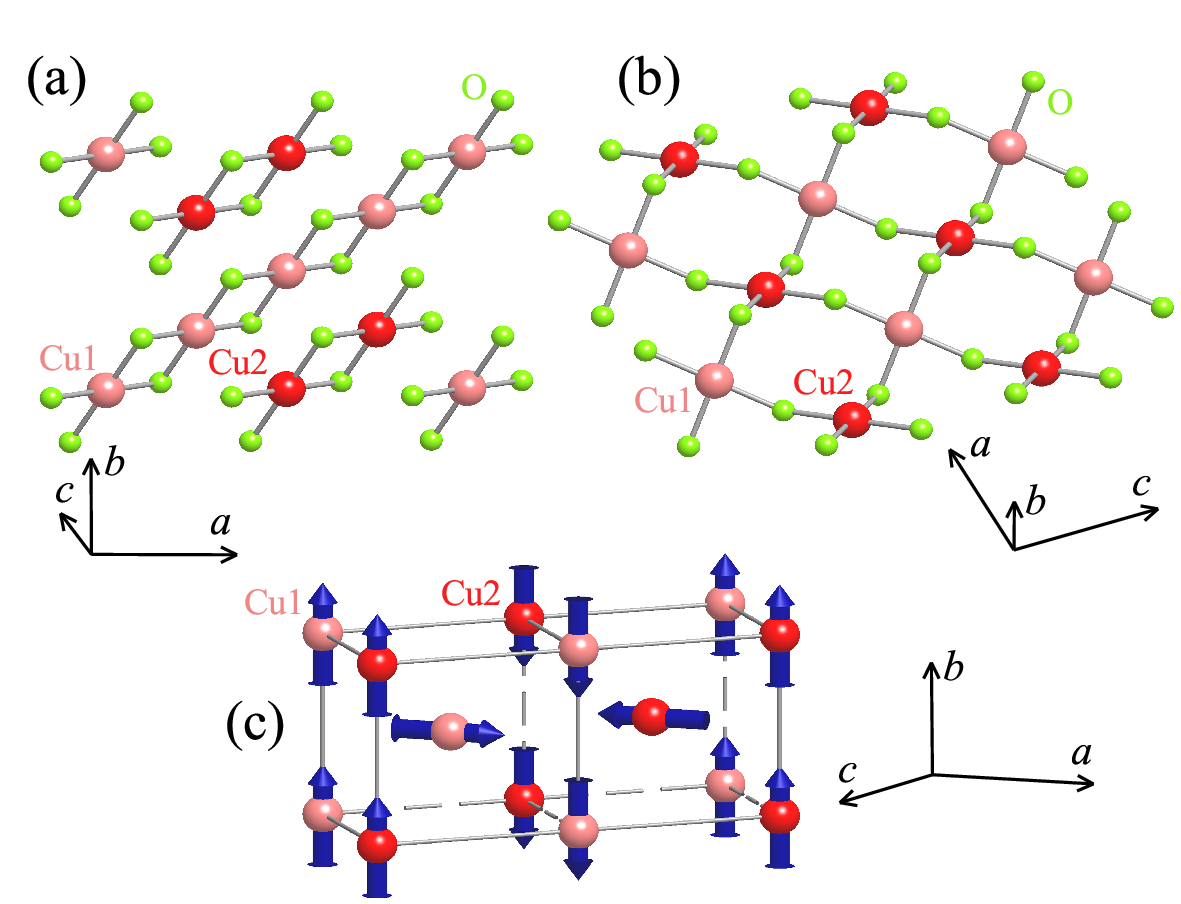}
\end{center}
\caption{
Crystal and magnetic structure of CuO. (a) The main structural patter in the $ab$ plane. (b) Fragment of the crystal structure in the $ac$ plane. (c) Noncollinear magnetic structure with $\vec{q} \approx (\frac{1}{2},0,-\frac{1}{2})$ leading to the multiferroic behavior. Two Cu sublattices are shown by different colors.}
\label{fig:CuO}
\end{figure}

 Cupric oxide (CuO) is the promising multiferroic material. It crystallizes in the monoclinic $C2/c$ structure~\cite{Kimura_CuO}. Similar to Cu$X_2$ ($X=$ Cl or Br), the main structural pattern of CuO is the centrosymmetric CuO$_2$ chains propagating in the $ab$ plane (Fig.~\ref{fig:CuO}a). The basic difference is that in CuO there are two Cu sublattices (Cu1 and Cu2 in Fig.~\ref{fig:CuO}), forming two distinct CuO$_2$ chains. Each sublattice can be transformed to itself by $\mathcal{I}$. However, different sublattices are not connected by $\mathcal{I}$. 
\noindent

\par At low temperatures, CuO is a collinear antiferromagnet. Nevertheless, when the temperature increases, it exhibits a transition to an incommensurate spin-spiral phase with $\vec{q}=(0.506,0,-0.483)$ (Fig.~\ref{fig:CuO}c) coexisting with the onset of multiferroic activity~\cite{Kimura_CuO,PRL2022}. The N\'eel temperature is relatively high ($T_{\rm N} = 230$ K) and can be further increased up to the room temperature by applying the hydrostatic pressure, as was suggested on the basis of first-principles electronic structure calculations~\cite{CuORT} and directly verified in experiments on the neutron diffraction~\cite{PRL2022}. Thus, CuO can become a room-temperature multiferroic, where the electric polarization is induced by the noncollinear magnetic order. 

\par The nominal spin of Cu$^{2+}$ in CuO is again $S=1/2$, suggesting that the single-ion anisotropy does not contribute to magnetic dependence of $\vec{P}$. Within centrosymmetric sublattices Cu1 and Cu2, the magnetic dependence of electric polarization can be described by generalized KNB mechanism. However, since Cu1 and Cu2 are not connected by $\mathcal{I}$, the coupling between these sublattices should involve the additional contribution given by Eq.~(\ref{eq:P12A}).

\subsection{\label{sec:Ba2CoGe2O7} Ba$_2$CoGe$_2$O$_7$ and Ba$_2$CuGe$_2$O$_7$}
\par
\begin{figure}[b]
\begin{center}
\includegraphics[width=8.6cm]{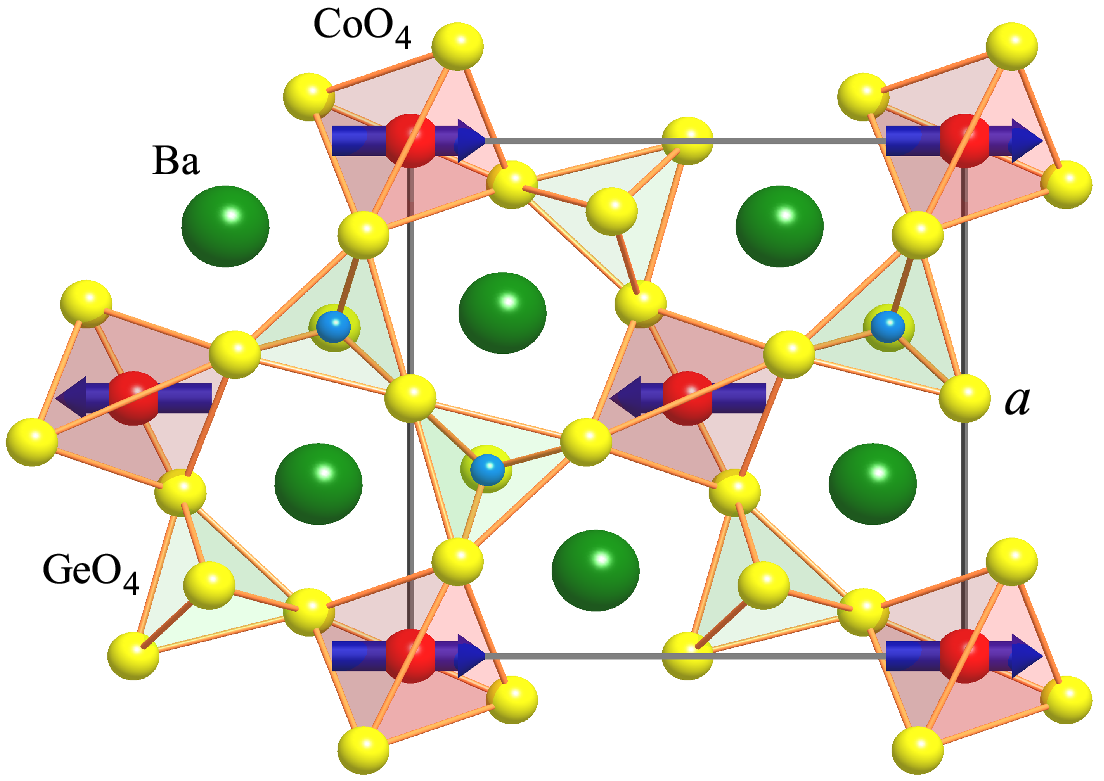}
\end{center}
\caption{
Crystal and magnetic structure of Ba$_2$CoGe$_2$O$_7$ in the tetragonal plane. $a$ is the lattice parameter.}
\label{fig:Ba2CoGe2O7}
\end{figure}

 Ba$_2$CoGe$_2$O$_7$ and Ba$_2$CuGe$_2$O$_7$ crystallize in the noncentrosymmetric tetragonal $P\overline{4}2_{1}m$ structure (see Fig.~\ref{fig:Ba2CoGe2O7})~\cite{Hutanu2011,Zheludev1996}.
\noindent
\noindent The crystal structure itself is non-polar due to the fourfold rotoinversion symmetry. Nevertheless, certain magnetic alignment can lower the symmetry and induce finite electric polarization. Besides the multiferroicity, the $P\overline{4}2_{1}m$ symmetry is responsible for a number of interesting effects, such as the coexistence of chiral magnetic ordering and weak ferromagnetism as well as the possibility of inducing the antiskyrmion textures~\cite{Bogdanov}. 

\par From the viewpoint of magnetism, the main difference between Ba$_2$CoGe$_2$O$_7$ and Ba$_2$CuGe$_2$O$_7$ is spin state of the magnetic ions: $S=3/2$ for Co$^{2+}$, while $S=1/2$ for Cu$^{2+}$. Thus, the single-ion contributions to the exchange energy and electric polarization are expected in Ba$_2$CoGe$_2$O$_7$, but not in Ba$_2$CuGe$_2$O$_7$. This has a dramatic effect on the magnetic properties of these compounds. The single-ion anisotropy in Ba$_2$CoGe$_2$O$_7$ traps the magnetic moments in the $xy$ plane and forces the nearly collinear AFM structure with $\vec{q}_{0}=(1,0,0)$ (see Fig.~\ref{fig:Ba2CoGe2O7}). The AFM coupling between the nearest spins is driven by isotropic exchange interactions. Since the single-ion anisotropy energy for $S=3/2$ in the tetragonal case does not depend on the azimuthal angle $\phi$~\cite{Skomski}, this magnetic structure can be easily rotated in the $xy$ plane by the magnetic field~\cite{Murakawa}. The DM interactions are permitted by the crystallographic symmetry. However, in Ba$_2$CoGe$_2$O$_7$ these interactions are overpowered by large single-ion anisotropy, resulting in collinear AFM alignment. In Ba$_2$CuGe$_2$O$_7$, the single-ion anisotropy vanishes so that the DM interactions become active, leading to formation of the incommensurate spin-spiral phase with $\vec{q} = \vec{q}_{0} + \delta \vec{q}$. The spin moments rotate in the plane specified by the vector $\vec{n}^{\perp} = (-\sin \phi, \cos \phi, 0)$, being perpendicular to the spin-rotation plane, and the spin-spiral deformation of the AFM $\vec{q}_{0}=(1,0,0)$ structure is given by $\delta \vec{q} = \delta q ( \sin \phi, \cos \phi, 0)$, where $\delta q$ depends on the ration of DM and isotropic exchange interactions between nearest neighbors in the $xy$ plane~\cite{PRB2020}. In the other words, the spin spiral propagates in the $xy$ plane, while the spins rotate in the plane formed by the $z$ axis and one of the axes in the $xy$ plane. The classical energy of the spin spiral does not depend on $\phi$. In order to find $\phi$, it is essential to consider the effects of quantum zero-point motion, following the procedure of Yildirim \textit{et al.}~\cite{Yildirim}, which tend to stabilize the magnetization parallel to one of the square diagonals in the $xy$ plane~\cite{PRB2020}.

\par The KNB mechanism is not activated in Ba$_2$CoGe$_2$O$_7$, which is the collinear antiferromagnet. The magnetic dependence of $\vec{P}$ in this case is almost solely given by the single-ion anisotropy: the rotation of magnetization in the $xy$ plane destroys the rotoinversion symmetry and yields finite $\vec{P}$~\cite{Murakawa,PRB2015b}. In the $S=\frac{1}{2}$ material Ba$_2$CuGe$_2$O$_7$, the single-ion anisotropy vanishes and the polarization is induced by the spin-spiral alignment via the generalized KNB mechanism. However, the spiral structure itself is driven by the DM interactions and emerges in the first order of the SO coupling. Therefore, the magnetic dependence of electric polarization is the second-order effect with respect to the SO coupling. Furthermore, all Cu-Cu bonds in Ba$_2$CuGe$_2$O$_7$ are noncentrosymmetric and besides KNB there can be other contributions to $\vec{P}$~\cite{PRB2020}.

\subsection{\label{sec:triangular} Triangular-lattice magnets}
\par
\begin{figure}[t]
\begin{center}
\includegraphics[width=8.6cm]{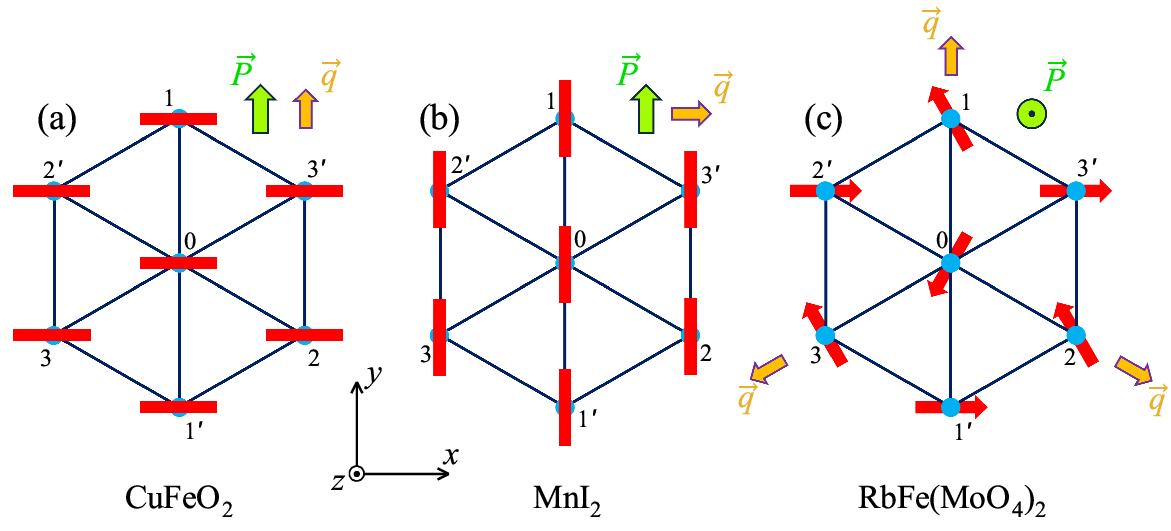}
\end{center}
\caption{
Different types of magnetic order on triangular lattices realized in CuFeO$_2$ (a), MnI$_2$ (b), and RbFe(MoO$_4$)$_2$ (c). $\vec{q}$ shows the propagation direction of the spin spiral. In the $120^{\circ}$ spin structure, realized in RbFe(MoO$_4$)$_2$, there are three equivalent directions of $\vec{q}$. $\vec{P}$ shows the corresponding direction of electric polarization. In (a) and (b), the magnetic moments are rotated in the planes formed by the $z$ axis and either $x$ (a) or $y$ (b) axis. These planes are indicated by thick red bars. In (c), the magnetic moments are in the $xy$ plane and indicated by the red arrows.}
\label{fig:triangular}
\end{figure}

 The triangular-lattice magnets CuFeO$_2$~\cite{Arima}, MnI$_2$~\cite{MnI2}, and RbFe(MoO$_4$)$_2$~\cite{RbFeMo2O8} have attracted a great deal of attention as examples of spin-spiral multiferroics, where the canonical KNB rules fail. Namely, the magnetic textures realized in CuFeO$_2$ and MnI$_2$ can be regarded as ``proper screw'', where the moments rotate in the plane perpendicular to the spin-spiral propagation $\vec{q}$ in the triangular plane (see Figs.~\ref{fig:triangular}a and \ref{fig:triangular}b).
\noindent
\noindent Therefore, the construction $\vec{q} \times [\boldsymbol{e}_{i} \times \boldsymbol{e}_{j}]$ should vanish. There are two possible configurations, where: (i) the rotation plane is perpendicular to the bonds of the triangular lattice (see Fig.~\ref{fig:triangular}a), which is realized in CuFeO$_2$~\cite{Arima}; (ii) the rotation plane is parallel to the bonds of the triangular lattice (Fig.~\ref{fig:triangular}b), which is realized in MnI$_2$, in low magnetic field~\cite{MnI2}. Furthermore, in MnI$_2$, the configuration shown in Fig.~\ref{fig:triangular}a can be also realized in the high magnetic field, applied perpendicular to the rotation plane~\cite{MnI2}. The proper screw magnetic order induces the electric polarization, unexpectedly from the viewpoint of KNB rules. The polarization remains parallel to the bonds for both types of magnetic structures shown in Figs.~\ref{fig:triangular}a and \ref{fig:triangular}b. In the $120^{\circ}$ spin structure, realized RbFe(MoO$_4$)$_2$, the magnetic moments rotate in the triangular plane. This magnetic structure can be described by three propagation vectors $\vec{q}$, which are parallel to the bonds (see Fig.~\ref{fig:triangular}c). Therefore, according to KNB rules, for each bond, the polarization is expected to be in the plane and perpendicular to this bond, so that there will be a cancellation of contributions stemming from three bonds around each magnetic site. However, the experiment situation is again different: there is a finite polarization and is perpendicular to the triangular plane~\cite{RbFeMo2O8}. 

\par These data spurred the search of new mechanisms of the magnetoelectric coupling, which would explain the properties of triangular-lattice magnets. This mechanism was regarded to be the spin-dependent metal-ligand hybridization~\cite{TokuraSekiNagaosa,Arima}, which is some sort of the single-ion effect as it does not depend on the relative alignment of spins on different magnetic sites. However, in all these materials, the magnetic atoms are located in the inversion centers. Therefore, the single-ion contributions to the electric polarization are forbidden by the symmetry.

\par Nevertheless, the behavior of CuFeO$_2$, MnI$_2$, and RbFe(MoO$_4$)$_2$ can be naturally understood by considering the generalized KNB mechanism for the magnetic bonds with relatively low symmetry~\cite{Xiang,KaplanMahanti}. Indeed, the bonds in the triangular plane are centrosymmetric. However, besides the spatial inversion, there is only twofold rotation in the case of CuFeO$_2$ and MnI$_2$, and no other symmetries at all in the case of RbFe(MoO$_4$)$_2$. 

\par In CuFeO$_2$, the spin-spiral propagation vector can be taken parallel to $y$: $\vec{q} = (0, q^{y}, 0)$ (see Fig.~\ref{fig:triangular}a). The spin-spiral plane is perpendicular to $\vec{q}$, both in CuFeO$_2$ and MnI$_2$. Therefore, we have $[\boldsymbol{e}_{0} \times \boldsymbol{e}_{1}] = (0, \sin q^{y}, 0)$ and $[\boldsymbol{e}_{0} \times \boldsymbol{e}_{2}] = [\boldsymbol{e}_{0} \times \boldsymbol{e}_{3}] = (0, -\sin \frac{q^{y}}{2}, 0)$, where $\vec{q}$ is in the units of inverse lattice parameter. The symmetry properties of $\vec{\boldsymbol{r}}_{01}$, $\vec{\boldsymbol{r}}_{02}$, and $\vec{\boldsymbol{r}}_{03}$ in the nearest bonds, which are connected by the threefold rotation, are considered in Supplemental Material~\cite{SM}. Then, it is straightforward to find that $\vec{P} = (0,P^{y},0)$, where $P^{y} \sim  r^{y,y} \sin q^{y} - \left( \frac{3}{2} r^{x,x} + \frac{1}{2} r^{y,y} \right) \sin \frac{q^{y}}{2}$. Furthermore, since $|r^{x,x}| \gg |r^{y,y}|$ ($r^{x,x}$ is of the in the SO coupling, while $r^{y,y}$ is only of the second order), one can write $P^{y} \sim  - \frac{3}{2} r^{x,x} \sin \frac{q^{y}}{2}$.

\par In MnI$_2$, $\vec{q}$ is parallel to $x$: $\vec{q} = (q^{x}, 0, 0)$ (see Fig.~\ref{fig:triangular}b). Then, $[\boldsymbol{e}_{0} \times \boldsymbol{e}_{1}]=0$ and this bond does not contribute to $\vec{P}$. For other two bonds we have $[\boldsymbol{e}_{0} \times \boldsymbol{e}_{2}]= -[\boldsymbol{e}_{0} \times \boldsymbol{e}_{3}] = (\sin \frac{\sqrt{3} q^{x}}{2}, 0, 0)$. Combining them with $\vec{\boldsymbol{r}}_{02}$ and $\vec{\boldsymbol{r}}_{03}$~\cite{SM}, it is again straightforward to find that $\vec{P} = (0,P^{y},0)$, where in the first order of the SO coupling $P^{y} \sim  -\frac{\sqrt{3}}{2} r^{x,x} \sin \frac{\sqrt{3} q^{x}}{2}$.

\par For the $120^{\circ}$ spin structure, realized in RbFe(MoO$_4$)$_2$ (see Fig.~\ref{fig:triangular}c), we have $[\boldsymbol{e}_{0} \times \boldsymbol{e}_{1}]= [\boldsymbol{e}_{0} \times \boldsymbol{e}_{2}]= [\boldsymbol{e}_{0} \times \boldsymbol{e}_{3}] = (0, 0, -\frac{\sqrt{3}}{2})$, resulting in $\vec{P} = (0,0,P^{z})$, where $P^{z} \sim -\frac{\sqrt{3}}{2} r^{z,z}$.  

\par Thus, the behavior of the triangular-lattice magnets CuFeO$_2$, MnI$_2$, and RbFe(MoO$_4$)$_2$ can be understood in terms of the generalized KNB mechanism. The two important points are: (i) the general form of the tensor $\vec{\boldsymbol{r}}$, which in the case of the low symmetry can have finite diagonal elements; (ii) The geometry of magnetic bonds in the triangular lattice. Since all bonds are centrosymmetric, no additional complications are expected.

\section{\label{sec:summary} Summary and outlook}
\par We have considered some basic principles of how the inversion symmetry can be broken by magnetic order, resulting in net electric polarization. The fundamental reason of such magnetic inversion symmetry breaking is the incompatibility of the FM and AFM order patterns, which coexist in multiferroic systems. If the FM order parameter is invariant under the spatial inversion, $\mathcal{I} \boldsymbol{F} = \boldsymbol{F}$, the AFM ones requires the additional time reversal, $\mathcal{IT} \boldsymbol{A} = \boldsymbol{A}$. Since $\mathcal{I}$ cannot coexist with $\mathcal{IT}$, the inversion symmetry is broken, that is manifested in the appearance of spontaneous polarization. The noncollinear alignment of spins is only one possible scenario where $\boldsymbol{F}$ coexists with $\boldsymbol{A}$ and there can be other possibilities besides the noncollinearity. For instance, the inversion symmetry breaking in the E phase of manganites is typically attributed to the exchange striction effects~\cite{CheongMostovoy,Khomskii2009,TokuraSekiNagaosa}. However, there is even more fundamental reason besides the exchange striction. This is again the incompatibility of the long-range AFM order in the Mn sublattice with the high-spin state of Mn$^{3+}$ ions located in the inversion centers. This readily explains the fact that, on the level of electronic structure calculations, the electric polarization can be readily obtained by imposing the E-type AFM order in the centrosymmetric orthorhombic structure without polar distortions~\cite{YMnO3,PRB2014,PRB2015,Picozzi2007}. Another example is GdFeO$_3$, where G-type antiferromagnetically ordered Gd spins obey the magnetoelectric $\mathcal{IT}$ symmetry, but the $\mathcal{IT}$ symmetry operation is incompatible with the magnetism of Fe atoms in the centrosymmetric positions~\cite{Yamaguchi}.  

\par This coexistence principle readily explain similarities and differences between the multiferroicity and magnetoelectricity, and how one can formally design a multiferroic materials starting from the magnetoelectric ones. The magnetoelectricity implies the AFM order, obeying the $\mathcal{IT}$ symmetry: $\mathcal{IT} \boldsymbol{A} = \boldsymbol{A}$. This symmetry operation can be broken by applying either electric or magnetic field, which induces simultaneously the ferroelectric polarization and ferromagnetic magnetization~\cite{DzyaloshinskiiME}. The multiferroicity is the phenomenon of the same type, but where the internal FM order parameter $\boldsymbol{F}$ plays the role of magnetic field, which breaks the $\mathcal{IT}$ symmetry. Loosely speaking, the simplest recipe for making multiferroic material starting from the magnetoelectric one is to add an magnetic atom to the inversion center.

\par The coexistence principle prescribes how the electric polarization $\vec{P}$ should depend on spins. In the centrosymmetric bond, this dependence can be only in the form of the antisymmetric coupling between the spins or the single-ion anisotropy. No other contributions are formally allowed. 

\par The microscopic picture behind these phenomenological principles can be derived by employing the modern theory of electric polarization~\cite{FE_theory1,FE_theory2,FE_theory3}. For Mott multiferroics, whose electronic and magnetic properties are controlled by large on-site Coulomb repulsion, the modern theory can be further reformulated as the superexchange theory for the electric polarization. Nowadays, the modern theory of electric polarization is implemented in most of the electronic structure codes based on the density functional theory so that $\vec{P}$ can be easily calculated without additional approximations, such as the superexchange theory. Nevertheless, if we are interested in \emph{interpretation} of experimental data or results of electronic structure calculations, such approximations become very useful as they provide a clear answer which microscopic invariant is responsible for finite $\vec{P}$ in the case of noncollinear spins or any other scenario of magnetic inversion symmetry breaking. Namely, for the noncollinear alignment of spins, the magnetoelectric coupling in the centrosymmetric bond is controlled by the spin-dependent part of the position operator in the basis of Kramers states, $\vec{\boldsymbol{r}}_{12}^{\phantom{0}}$.

\par The form of $\vec{\boldsymbol{r}}_{12}^{\phantom{0}}$ is not universal and depends on the symmetry of the bond. The symmetry properties of $\vec{\boldsymbol{r}}_{12}^{\phantom{0}}$ can be derived by considering the symmetry of the Kramers states. Particularly, the commonly used KNB rule $\vec{P} \propto \vec{\epsilon}_{21} \times [\boldsymbol{e}_{1} \times \boldsymbol{e}_{2}]$ is justified only for relatively high rotational symmetry of the bond (threefold or higher). Considering the applicability of the KNB rules to the real multiferroic materials, the primary question is whether the magnetic bonds are centrosymmetric or not. If all bonds in the crystal are centrosymmetric, the KNB mechanism itself is applicable, though the rules connecting the directions of spins with the electric polarization can be modified to take into account the proper symmetry of the bond, which is typically rather low~\cite{Xiang,KaplanMahanti}. Yet, the magnetoelectric coupling in this case remains antisymmetric with respect to the permutation of $\boldsymbol{e}_{1}$ and $\boldsymbol{e}_{2}$. However, if the bonds are noncentrosymmetric, there can be many complications, related to the existence of other contributions to $\vec{P}$, of isotropic, antisymmetric, and isotropic symmetric types. 

\par From the viewpoint of symmetry of individual bonds, there are not so many materials where the conventional KNB rule can be realized as it is, without modifications. One of such examples is magnetoelectric Cr$_2$O$_3$~\cite{DzyaloshinskiiME}, those Cr-Cr bonds have threefold rotational symmetry. Nevertheless, another important factor is that, in solid, the behavior of polarization is not necessarily related to the properties of a single bond: there are typically several bonds contributing to $\vec{P}$, these bonds can have different orientation in space, so that the observable polarization is an averaged property, which can have higher symmetry.

\par The single-ion anisotropy part of $\vec{P}$ is also subjected to several constraints. It is expected to vanish for the spin $\frac{1}{2}$ and if magnetic ions are located in the inversion centers, which impose severe restrictions on possible involvement of this mechanism in the multiferroic behavior of real materials. Particularly, considering the properties of proper screw multiferroics, where the conventional KNB mechanism was expected to fail, prompting alternative scenarios based on the single-ion anisotropy~\cite{Arima}, most of these magnets appeared to be either spin-$\frac{1}{2}$ or to have the magnetic ions in the centrosymmetric positions. Therefore, the single-ion anisotropy mechanism is not applicable. Instead, the properties of proper screw multiferroics can be described by the generalized KNB mechanism, taking into consideration the proper symmetry of magnetic bonds. 

\par Thus, it is already more than twenty years after discovery of multiferroicity - a seemingly exotic phenomenon of breaking the inversion symmetry by a magnetic order, resulting in ferroelectric polarization. The main purpose of this review was to show that, in principle, there is nothing exotic: the magnetic inversion symmetry breaking obey very clear principles, which can be readily established on phenomenological and microscopical levels. The KNB theory was the first microscopic theory of electric polarization in noncollinear magnets. This theory can be systematically understood and extended on the grounds of modern theory of electric polarization in periodic systems.

\section*{Acknowledgement}
\par I am grateful to Akihiro Tanaka for useful suggestions, help, and constant encouragement in the process of working on this review. MANA is supported by World Premier International Research Center Initiative (WPI), MEXT, Japan.

\end{document}